\documentclass[11.5pt,preprint]{aastex}

\usepackage{graphicx} 
\usepackage{lscape}
\usepackage{natbib}
\newcommand\simlt{\hspace{0.3em}\raisebox{0.4ex}{$<$}\hspace{-0.75em}\raisebox{-.7ex}{$\sim$}\hspace{0.3em}}

\slugcomment{Accepted for Publication in the Astronomical Journal}

\shorttitle{Near Infrared Imaging Survey of Bok Globules}
\shortauthors{Kandori et al.}

\begin{document}

\title{Near Infrared Imaging Survey of Bok Globules: Density Structure}

\author{Ryo Kandori\altaffilmark{}} 
\affil{Department of Astronomical Science, Graduate University for Advanced Studies (Sokendai), 2-21-1 Osawa, Mitaka, Tokyo 181-8588, Japan; Present address: Optical and Infrared Astronomy Division, National Astronomical Observatory of Japan, 2-21-1 Osawa, Mitaka, Tokyo 181-8588, Japan (kandori@optik.mtk.nao.ac.jp)}

\author{Yasushi Nakajima\altaffilmark{}, Motohide Tamura\altaffilmark{}, Ken'ichi Tatematsu\altaffilmark{}} 
\affil{National Astronomical Observatory of Japan, 2-21-1 Osawa, Mitaka, Tokyo 181-8588, Japan (yas@optik.mtk.nao.ac.jp, hide@optik.mtk.nao.ac.jp, k.tatematsu@nao.ac.jp)}

\author{Yuri Aikawa\altaffilmark{}} 
\affil{Department of Earth and Planetary Sciences, Kobe University, Kobe 657-8501, Japan (aikawa@kobe-u.ac.jp)}

\author{Takahiro Naoi\altaffilmark{}} 
\affil{Department of Earth and Planetary Science, The University of Tokyo, 7-3-1 Hongo, Bunkyo-ku, Tokyo 113-0033, Japan (naoi@subaru.naoj.org)}

\author{Koji Sugitani\altaffilmark{}} 
\affil{Institute of Natural Sciences, Nagoya City University, Mizuho-ku, Nagoya 467-8501, Japan (sugitani@nsc.nagoya-cu.ac.jp)}

\author{Hidehiko Nakaya\altaffilmark{}} 
\affil{Subaru Telescope, National Astronomical Observatory of Japan, 650 North A{\' o}hoku Place, Hilo, HI 96720, USA (nakaya@subaru.naoj.org)}

\author{Takahiro Nagayama\altaffilmark{}, Tetsuya Nagata\altaffilmark{}} 
\affil{Department of Astronomy, Kyoto University, Kyoto 606-8502, Japan
(nagayama@kusastro.kyoto-u.ac.jp, nagata@kusastro.kyoto-u.ac.jp)}

\author{Mikio Kurita\altaffilmark{}, Daisuke Kato\altaffilmark{}, Chie Nagashima\altaffilmark{}, and Shuji Sato\altaffilmark{}} 
\affil{Department of Astrophysics, Nagoya University, Chikusa-ku, Nagoya 464-8602, Japan (mikio@z.phys.nagoya-u.ac.jp, kato@z.phys.nagoya-u.ac.jp, chie@z.phys.nagoya-u.ac.jp, ssato@z.phys.nagoya-u.ac.jp)}

\begin{abstract}
On the basis of near-infrared imaging observations, we derived visual extinction ($A_{V}$) distribution toward ten Bok globules through measurements of both the color excess ($E_{H-K}$) and the stellar density at $J$, $H$, and $K_{\rm {s}}$ (star count). Radial column density profiles for each globule were analyzed with the Bonnor-Ebert sphere model. Using the data of our ten globules and four globules in the literature, we investigated the stability of globules on the basis of $\xi {}_{{\rm max}}$, which characterizes the Bonnor-Ebert sphere as well as the stability of the equilibrium state against the gravitational collapse. We found that more than half of starless globules are located near the critical state ($\xi {}_{{\rm max}} = 6.5 \pm 2$). Thus, we suggest that a nearly critical Bonnor-Ebert sphere characterizes the typical density structure of starless globules. Remaining starless globules show clearly unstable states ($\xi {}_{{\rm max}} > 10$). Since unstable equilibrium states are not long maintained, we expect that these globules are on the way to gravitational collapse or that they are stabilized by non-thermal support. It was also found that all the star-forming globules show unstable solutions of $\xi {}_{\rm max}>10$, which is consistent with the fact that they have started gravitational collapse. We investigated the evolution of a collapsing gas sphere whose initial condition is a nearly critical Bonnor-Ebert sphere. We found that the column density profiles of the collapsing sphere mimic those of the static Bonnor-Ebert spheres in unstable equilibrium. The collapsing gas sphere resembles marginally unstable Bonnor-Ebert spheres for a long time. We found that the frequency distribution of $\xi {}_{\rm max}$ for the observed starless globules is consistent with that from model calculations of the collapsing sphere. In addition to the near-infrared observations, we carried out radio molecular line observations (C$^{18}$O and N$_{2}$H$^{\rm +}$) toward the same ten globules. We confirmed that most of the globules are dominated by thermal support. The line width of each globule was used to estimate the cloud temperature including the contribution from turbulence, with which we estimated the distance to the globules from the Bonnor-Ebert model fitting. 
\end{abstract}


\section{Introduction} 
Probing the physical evolution of dense cores toward the onset of star formation is of great importance in understanding the nature of the star formation process. The star formation process as well as the characteristics of forming stars are probably mostly determined by the properties of the nursing cores, such as size, mass, temperature, turbulence, and density structure, right before the gravitational collapse. In order to clarify the initial condition of star formation, it is important to provide a number of dense core samples with well defined physical properties.  
\par
Bok globules (i.e., isolated molecular cloud cores) are ideal sites for studying isolated star formation processes because of their simple shape and isolated geometry from neighboring clouds. These characteristics of globules enable us to model their internal structure (e.g., density). For these reasons, detailed observations of some well-known globules have been done. Alves et al. (2001) reported a high-resolution dust-extinction ($A_{V}$) study of the starless globule Barnard 68, based on near-infrared observations. They presented a radial column density profile of the globule which is well fitted by the Bonnor-Ebert sphere model (Bonnor 1956; Ebert 1955). Because the density structure is well determined, molecular abundances in Barnard 68 have been studied by comparing dust extinction with molecular distribution (Bergin et al. 2002; Hotzel et al. 2002a, 2002b; Di Francesco et al. 2002; Lai et al. 2003). 
\par
An unbiased survey of both starless and star-forming globules is important to provide essential information on the evolution of density, velocity, and chemical structure of globules as well as their relative lifetime. Statistical studies of globules (dense cores) have been conducted based on the observations of thermal dust emission (e.g., Visser et al. 2001, 2002), dust extinction in the mid-infrared (Bacmann et al. 2000), molecular line emission (e.g., Launhardt et al. 1998; Caselli et al. 2002), and a combination of these (e.g., Tafalla et al. 2002; dust and molecular line emissions). 
However, no statistical studies of the density structure of globules based on the near-infrared extinction have been carried out. Recently, the dust temperature gradient in cold starless cores (e.g., Evans et al. 2001; Ward-Thompson, Andre, \& Kirk 2002) has been pointed out, and this may systematically affect the column density distribution derived by assuming a constant dust temperature. 
The measurements of extinction in the mid-infrared surface photometry (e.g., Bacmann et al. 2000) can also be affected by dust temperature distributions, because both absorption and emission from the line-of-sight medium should be taken into account at these wavelengths. 
As a column density tracer, near-infrared extinction measurement is one of the most straightforward methods with no ambiguity caused by dust temperature distribution. Near-infrared extinction measurements can trace dust column densities at low-density outer regions of the globules, where (sub)millimeter dust emission is hard to detect. 
\par
The aim of our near-infrared ($J$, $H$, and  $K_{s}$) imaging survey of Bok globules is to reveal the evolution of radial density structures of globules from the starless to protostellar phases as well as to derive their fundamental physical parameters (e.g., size, temperature, mass, and external pressure) on the basis of the measurements of dust extinction. We investigated the radial $A_{V}$ distribution for each globule using the Bonnor-Ebert sphere model. The solutions of the Bonnor-Ebert sphere can be characterized by the dimensionless radial parameter $\xi {}_{{\rm max}}$, which describes the stability of the gas sphere against gravitational collapse, and the critical state is achieved at $\xi {}_{{\rm max}}$=6.5. 
By combining our results for ten Bok globules with previous reports on four sources, we investigate the difference in physical properties, in particular the $\xi {}_{{\rm max}}$ value, between the globules in the starless phase and those in the star-forming (protostellar or gravitational collase) phase, which is likely to be related to the difference in evolutionary stages of the globules. In order to investigate the velocity structure of globules, we observed the same globules using the 45 m radio telescope of Nobeyama Radio Observatory with the C${}^{18}$O ($J=1\rightarrow 0$) and/or N${}_{2}$H${}^{+}$ ($J=1\rightarrow 0$) molecular lines. 
\par
The procedures of our observations and data reduction are described in $\S$ 2. The distribution of dust extinction ($A_V$) for each globule is obtained in order to derive the radial $A_V$ profile and to fit it by the Bonnor-Ebert sphere model ($\S$ 3 and 4). We investigate the difference in the physical properties between the starless and star-forming globules and discuss implications for the evolution of globules in $\S$ 5. Our conclusions are summarized in $\S$ 6. 

\section{Observations and Data Reduction}
\subsection{Source Selection}
We observed ten Bok globules, CB 87, CB 110, CB 131, CB 134, CB 161, CB 184, CB 188, FeSt 1-457, Lynds 495, and Lynds 498. These were selected from a unified catalog of dark clouds (Dutra \& Bica 2002; compilation of 21 dark cloud catalogs) with the following criteria: (1) globules whose angular sizes are less than $\sim$5$'$, (2) those having simple shapes (circular or elliptical), (3) those well isolated from neighboring clouds, (4) those located in the solar neighborhood (distance of less than $\sim$500 pc), and (5) those observable from both hemispheres. 
Our selection criteria of (1) to (4) are similar to those used in the CB catalog (Clemens \& Barvainis 1988), which lists 248 small, isolated, optically selected dark clouds. Typical angular size and ellipticity of dark clouds in the CB catalog are $\sim 3'$ and $\sim 2$, respectively (see, Figures 1 and 2 of Clemens \& Barvainis 1988). In the CB catalog, the fraction of large (e.g., $\gg 5'$) and/or highly elliptical (e.g., ellipticity $\gg 2$) clouds with isolated geometry is small. 
The properties of the observed globules are summarized in Table 1. Column 1 lists the name of the globule, Columns 2 and 3 list the equatorial coordinates (J2000) of the center of the target field. 
Column 4 gives the distance to the globule from the catalog (Dutra \& Bica 2002). The distance to the nearby and compact globules is difficult to determine. 
The lack of foreground stars toward globules makes spectro-photometric distance measurements impossible, and the distance estimation from LSR velocity (kinematic distance) is not reliable for nearby clouds. Thus, the listed distances are only values assumed by associating observed globules with neighboring larger molecular clouds whose distance is known. 
Column 5 represents the association of IRAS point sources within the optical boundary of each globule. 
We classified globules into starless and star-forming clouds based on the association of IRAS point sources\footnote{Recently, a low luminosity protostar was discovered toward the globule Lynds 1014, which is known to accompany no IRAS point sources, using the Spitzer Space Telescope (Young et al. 2004). Though Spitzer data are useful to clarify whether globules contain protostars or not, Spitzer data is available only for one source, CB 131, among the $\lq \lq$IRAS-less" globules in Table 1. We briefly checked mid-infrared data of the Spitzer telescope toward CB 131 and found that there is no protostar candidates near the center of the globule. Though a source is located near the globule on the 24 $\mu$m image, it is not known whether the source is a protostar associated with CB 131.}. 
\rm
A Class I protostar (IRAS 19179+1129) is associated with CB 188 (Launhardt 1996; Yun \& Clemens 1992), whereas no evidence of star formation is shown for the other nine globules. For CB 184, a Class II source (IRAS 19116+1623) is located near the globule (Launhardt 1996), but its physical association with the globule is not known. Since star-forming activity is not taking place around the center of CB 184, we will treat CB 184 as a starless globule in this paper. 
The Bonnor-Ebert model fitting studies of globules based on near-infrared extinction have been reported in four sources in the literature: Barnard 68 (Alves et al. 2001), Barnard 335 (Harvey et al. 2001), Coalsack Globule II (Lada et al. 2004; Racca et al. 2002), and Lynds 694-2 (Harvey et al. 2003). A Class 0 protostar (IRAS 19345+0727) is associated with Barnard 335. The other globules have no IRAS sources, while Lynds 694-2 shows strong evidence of gas inward motion (Lee, Myers, \& Tafalla 1999, 2001; Lee, Myers, \& Plume 2004). Thus, we will classify Lynds 694-2 as a star-forming globule in this paper. There are two recent reports on the density structure of Coalsack Globule II based on similar analyses. We mainly refer to the observations of Lada et al. (2004) because their near-infrared observations are more sensitive than those of Racca et al. (2002). 

\subsection{Near-Infrared Data}
We carried out near-infrared ($J$, $H$, and  $K_{s}$) imaging observations of Bok globules using the infrared camera SIRIUS (Simultaneous three-color InfraRed Imager for Unbiased Surveys; Nagayama et al. 2003; Nagashima et al. 1999) on the 1.4 m telescope, IRSF (InfraRed Survey Facility), at the South African Astronomical Observatory (SAAO). Observations were made over two periods from 2002 July 30 to August 12 and from 2003 June 24 to July 14. SIRIUS has three 1024$\times$1024 pixel HgCdTe infrared detectors (HAWAII array), which enables us to take $J$, $H$, and  $K_{s}$ images simultaneously. IRSF/SIRIUS can image a large FOV of $\sim$7.$\hspace{-3pt}'$7$ \times$7.$\hspace{-3pt}'$7 with a scale of 0.$\hspace{-3pt}''$45/pixel, and the limiting magnitudes (S/N=10) for a typical 10-minute exposure time reach 18.9, 18.3, and 17.3 mag at $J$, $H$, and  $K_{s}$, respectively. Thus, this survey is much deeper than previous extensive surveys (e.g., 2MASS, DENIS). 
For each globule, we performed a 15-minute integration with a set of dithered 20 s or 30 s exposures except CB 110 and FeSt 1-457, whose integration times are 45 and 60 minutes, respectively. The dithered spatial interval was 30$''$ or 40$''$. Since the apparent diameter of each target globule is less than 5$'$, we covered its entire extent at one time. The typical seeing during the observations was 0.$\hspace{-3pt}''$9$ -$1.$\hspace{-3pt}''$4 (FWHM), corresponding to 2$-$3 pixels. To correct detector flatness and pixel-to-pixel variation, we obtained twilight flat and dark frames at the beginning and end of the nights. We observed some sets of near-infrared standard stars listed by Persson et al. (1998) during the observations for photometric calibration. 
\par
We processed the observed data in a standard way of near-infrared image reduction (i.e., dark subtraction, flat-field correction, and median sky subtraction) with the IRAF\footnote{IRAF is distributed by the National Optical Astronomy Observatories, which are operated by the Association of Universities for Research in Astronomy, Inc., under cooperative agreement with the National Science Foundation.} (Image Reduction and Analysis Facility) and the IDL (Interactive Data Language, by Research Systems, Inc.) routines. After subtraction of averaged dark frame, each object frame was divided by normalized flat frames. 
For correcting thermal emission pattern and the fringe pattern from the atomospheric OH emission, we constructed a median sky frame to be subtracted from the object frame. 
Since all the observed targets are located in the very crowded field of the Milky Way, we did not adopt the self-sky, with which the residual signal of (bright) stars cannot be excluded on the resulting self-sky frame. Therefore, we constructed the sky frame for each band by taking the median of dithered frames toward the different several (four to eight) objects observed in the same night, which resulted in a sky frame with no stellar residuals. This $\lq \lq$average" median sky does not reflect the actual sky for each object because the median was taken across several object frames. 
We compared the $\lq \lq$average" median sky frame with the self-sky frame for each object in order to correct their different sky levels, and the final (level-matched) median sky frames were made for each object. 
Though the thermal emission pattern is adequately corrected in the final median sky frames, the OH fringe pattern is time-averaged. 
We confirmed that the effect of the time-variation of the OH fringe pattern is within acceptable levels by checking the sky noise level of the final combined object frames. 
After the sky subtraction and frame-registration, we combined all the dithered images for each object with a 3$\sigma$ clipping. We show three-color ($JHK_s$) composite images of globules in Figure 1. 
\par
For the source detection and photometry, we applied the IRAF/DAOPHOT package (Stetson 1987) to the images. We measured the instrumental magnitudes of the detected stars having a peak intensity greater than 4.0 $\sigma$ noise level above the sky background by fitting a point-spread function (PSF) and subtracting it from the image. We iteratively performed stellar detections and psf-fitting photometries to the source-subtracted image in order to collect overlooked sources in the previous procedure. We manually checked the results of the photometries and removed some false detections by visual inspection. We summarize the results of stellar detection in Table 2. For calculating the plate solution, we matched the pixel coordinates of a number of detected stars with the celestial coordinates of their counterparts in the 2MASS point source catalog (Cutri et al. 2003) and obtained the transformation coefficient using the IRAF/IMCOORDS package, which resulted in an acceptable positional error of less than 0.$\hspace{-3pt}''$5 (rms). 
All the stars detected in our observations were cataloged separately with respect to the filters ($JHK_{s}$). These stars of single-band detections were used to generate an extinction map at each band with a star count method. We applied the following equations for the color conversion from the IRSF to the standard CTIO/CIT photometric system: 
\begin{eqnarray}
{(J-H)}_{\rm CIT} &=& 0.980 {\rm \times} {(J-H)}_{\rm IRSF} + 0.029\ \ \ {\rm for}  \ {(J-H)}_{\rm IRSF} \ge 0.35, \\
{(J-H)}_{\rm CIT} &=& 1.047 {\rm \times} {(J-H)}_{\rm IRSF}   \ \ \ \ \ \ \ \ \ \ \ \ \ \ {\rm for}  \ {(J-H)}_{\rm IRSF} < 0.35, \\
{(H-K)}_{\rm CIT} &=& 0.912 {\rm \times} {(H-{K}_{S})}_{\rm IRSF}
\end{eqnarray}
These were derived from measurements of red standard stars (see, Nakajima et al. 2004 for details). 

\subsection{Molecular Line Data}
We have carried out molecular line observations toward ten globules using the 45 m radio telescope of Nobeyama Radio Observatory over four observing periods (2003 March 10-19, 2003 April 8-10, 2004 January 9-12, and 2004 February 3-10). The observed objects are the same as those observed by IRSF/SIRIUS. 
The receiver employed was the 25-element focal-plane array receiver BEARS consisting of double-sideband SIS mixers (Sunada et al. 2000). All the target globules in Table 1 are mapped in position-switching mode in C${}^{18}$O ($J=1\rightarrow 0$) at 109.782173 GHz (JPL catalog; see, http://spec.jpl.nasa.gov/) and/or N${}_{2}$H${}^{+}$ ($J=1\rightarrow 0$) at 93.1737767 GHz (Caselli, Myers \& Thaddeus  1995) on a 13.$\hspace{-3pt}''$7 grid spacing. 
The half-power beamwidth for each element beam was estimated to be $\sim$17.$\hspace{-3pt}''$8 at 93 GHz and $\sim$15.$\hspace{-3pt}''$1 at 109 GHz. The map coverage ($\sim$4$'\times$4$'$) is sufficiently large compared with the extinction feature of globules. As a back end, we used a digital autocorrelator with a 37.8 kHz frequency resolution ($\sim$0.12 kms${}^{-1}$ at 93 GHz, and $\sim$0.10 kms${}^{-1}$ at 109 GHz). 
The intensity was calibrated using the standard chopper-wheel method, and is expressed in terms of the corrected antenna temperature $T_{A}^{*}$. To correct for the sideband ratio, we calibrated $T_{A}^{*}$ using the intensity obtained with the single-sideband receiver S100. 
The scaling factor from $T_{A}^{*}$ of BEARS to that of S100 for N${}_{2}$H${}^{+}$ ($J=1\rightarrow 0$) was measured by ourselves, and that for C${}^{18}$O ($J=1\rightarrow 0$) was provided by the observatory. A typical value of the scaling factor for each BEARS element was $\sim 1.5$. We estimated the main beam efficiency of S100 at 93 and 109 GHz to be 0.515 and 0.442, respectively, by interpolating values at 86, 100, and 115 GHz provided by the observatory. The pointing accuracy was typically better than 7$''$ by observing SiO maser sources at 43 GHz every 1-1.5 hours during the observations. 

\section{Derivation of $A_{V}$ and Molecular Line Parameters}
$A_{V}$ measurement is one of the most straightforward ways to probe the density structure of dark clouds. 
There are two standard ways to do extinction mapping; one is the star count method (e.g., Dickman 1978; Cambr\'{e}sy 1999; Dobashi et al. 2005) based on the measurements of the stellar density, and the other is the near-infrared color excess (NICE) method (Lada et al. 1994) based on measurements of the stellar color excess (e.g., $E_{H-K}$). 
By considering the stellar density in $H$ and $K_s$ at the densest region of globules, we decided which method is better for each target globule. 
We analyzed CB 131, CB 134 and CB 188 with the star count method and analyzed the other globules with the NICE method. 

\subsection{Star Count}
On the basis of the stars detected in the $J$, $H$, and  $K_{s}$ bands, we derived extinction maps ($A_{J}$, $A_{H}$, and $A_{K_{s}}$) for CB 131, CB 134, and CB 188 using the star count method (see, e.g., Kandori et al. 2003). 
A logarithmic cumulative stellar density $\log N$ measured at the magnitude $m_{\lambda,0}$ can be converted to extinction $A_{\lambda}$ as 
\begin{equation}
{A}_{ \lambda }\left({ \Delta \alpha , \Delta \delta ,{m}_{ \lambda ,0}}\right)={m}_{ \lambda ,0}-{f}^{-1}\left[{\log N( \Delta \alpha , \Delta \delta ,{m}_{ \lambda ,0})}\right]
\end{equation}
where $\Delta \alpha$ and $\Delta \delta$ are the offset positions from the equatorial coordinate of image center, and $f^{-1}$ is the inverse function of $f= \log N_{\rm ref} (m_{\lambda})$, which is the logarithmic cumulative luminosity function (i.e., Wolf diagram; Wolf 1923) constructed in the extinction-free reference field (see Fig. 2). Though $f$ is often assumed to be linear as $f=a+bm_{\lambda}$, we fitted it with a 4th-order polynomial because the function shows a slightly curved shape. 
\par
To measure stellar density distribution, $\log N(\Delta \alpha,\Delta \delta,{m}_{\lambda,0})$, we arranged circular cells (typically $\sim$30$''$ in diameter) separated by 9$''$ intervals on each observed field ($\sim$7$' \times $7$'$ area), and counted the number of stars falling in the cells up to the threshold magnitude $m_{\lambda,0}$ (see Table 2 for the definition) which is well above the limiting magnitudes. The diameter of circular cells depends on the stellar density in each cloud field. We smoothed the stellar density map with a Gaussian filter (FWHM$=$13.$\hspace{-3pt}''$5) to reduce noise. 
In order to derive $A_{J}$, $A_{H}$, and $A_{K_{s}}$, we numerically calculated $f^{-1}$ in equation (4) after substituting the actually measured stellar density $\log N(\Delta \alpha,\Delta \delta,{m}_{\lambda,0})$ into $f^{-1}$. 
In the resulting extinction maps, $A_J$ tends to be saturated at the densest region of the cloud, but sufficiently reveals the diffuse extinction feature around the cloud, and $A_{K_{s}}$ has the opposite tendency. 
In order to relieve saturation, we made a composite $A_{V}$ map by combining the $A_{J}$, $A_{H}$, and $A_{K_{s}}$ data with the following equation: 
\begin{eqnarray}
{A}_{V}&=& \alpha {R}_{K_{s}V}{ A}_{ K_{s}} +(1- \alpha ){R}_{HV}{A}_{H}\ \ \ {\rm for}\ {A}_{H} \ge { A}_{ H}^{\rm max} /2 \\
{A}_{V}&=& \beta {R}_{HV}{ A}_{ H} +(1- \beta ){R}_{JV}{A}_{J}\ \ \ \ \ {\rm for}\ {A}_{H}<{A}_{H}^{\rm max}/2
\end{eqnarray}
where $\alpha$ and $\beta$ are the weighting parameters defined as $A_{H}$/$A_{H}^{\rm max}$ and $A_{H}$/($A_{H}^{\rm max}/2$) where $A_{H}^{\rm max}$ denotes the maximum value of $A_{H}$ in each analyzing field. 
The ${R}_{\lambda V}$ denotes the standard conversion factor from $A_{\lambda}$ to $A_{V}$ in Rieke \& Lebofsky (1985), which is 3.55 and 5.71 for $J$ and $H$, respectively. 
For $K_{s}$, we assumed $A_{V}=9.44A_{K_{s}}$ on the basis of equation (3) and the conversion factors in Rieke \& Lebofsky (1985). 
These conversion factors depend on the assumption of dust optical properties, in particular $R_{V}$. We discuss systematic uncertainties in the derivation of $A_{V}$ in the Appendix. 
The resulting $A_{V}$ maps for CB 131, CB 134, and CB 188 are shown in Figure 2. 
Typical uncertainty of $A_{V}$ in the reference field ($A_{V} \sim 0$ mag) is estimated to be $\sim$0.9 mag. 

\subsection{Near-Infrared Color Excess}
On the basis of the stars detected both in the $H$, and $K_{s}$ bands, we derived extinction ($A_{V}$) maps for CB 87, CB 110, CB 161, CB 184, FeSt 1-457, Lynds 495, and Lynds 498 using the NICE method. 
If we assume that the population of stars toward the observed field is invariable, the mean stellar color in the reference field $<H-K>_{\rm ref}$ can be used as the zero point of color excess. The color excess distribution toward globules can be derived by subtracting $<H-K>_{\rm ref}$ from the observed $H-K$ of stars. 
Since the extent of each analyzed field is relatively small ($\sim 7'\times7'$), our assumption on invariant stellar population seems plausible. To measure color excess distribution, we arranged circular cells ($\sim$30$''$ diameter) spaced at 9$''$ (the same alignment as used in the star count analysis) on each observed field. We calculated the mean stellar color in each cell and derived their color excess as 
\begin{equation}
{E}_{H-K}\left({ \Delta \alpha , \Delta \delta }\right)=\left[{ \sum\limits_{ i=1}^{ N} {\frac{ (H-K{)}_{i}}{N}}}\right]\left({ \Delta \alpha , \Delta \delta }\right)-{\left\langle{H-K}\right\rangle}_{\rm ref}, 
\end{equation}
where $(H-K){}_{i}$ is the color index of the $i$-th star in a cell, and $N$ is the number of stars falling in a cell. $\langle H-K\rangle {}_{\rm ref}$ is the mean color of stars in the reference field (see, Fig. 2). 
We converted the $H-K$ color excess to $A_{V}$ using the reddening law of Rieke \& Lebofsky (1985) as $A_{V}=15.9\times E_{H-K}$. 
We then smoothed the resulting $A_{V}$ map in the same manner as used in the star count analysis. The resulting $A_{V}$ maps are shown in Figure 2. 
Typical uncertainty of $A_{V}$ in the reference field ($A_{V} \sim0$ mag) is estimated to be $\sim$0.6 mag. 

\subsection{Molecular Line Parameters}
We derived the observed molecular line parameters, LSR velocity, line width, and optical depth, for each globule. We observed N${}_{2}$H${}^{+}$ ($J=1\rightarrow 0$) for FeSt 1-457 and C${}^{18}$O ($J=1\rightarrow 0$) for the other globules. We show the spectrum observed at the center of each globule in Figure 3. For the N${}_{2}$H${}^{+}$ spectrum fitting we took into account their hyperfine structure and line optical depth effect (using the same procedure as described in Tatematsu et al. 2004). The intrinsic relative intensity and rest frequency of hyperfine components are taken from Tin\'{e} et al. (2000) and Caselli et al. (1995), respectively. 
\par
The observed line width $\Delta {V}_{\rm obs}$ of the globule ranges from $\sim 0.2$ to $\sim 0.5$ km s${}^{-1}$ (Table 3). The kinetic temperature $T_{\rm k}$ of low mass dense cores is known to be $\sim 10$ K (e.g., Benson \& Myers 1989). Thus, it is natural to attribute the excess in the observed line width with respect to the thermal line width ($0.13$ km s${}^{-1}$ for C${}^{18}$O and N${}_{2}$H${}^{+}$) to the contribution from non-thermal turbulence in globules. We estimated non-thermal line width $\Delta {V}_{\rm NT}$ of each globule by assuming that $T_{\rm k}$ is equal to 10 K. $\Delta {V}_{\rm NT}$ can be obtained from $\Delta {V}_{\rm NT}^{2} = \Delta {V}_{\rm obs}^{2} - 8 {\rm ln} 2k{T}_{\rm k}/{m}_{\rm obs}$, where $m_{\rm obs}$ is the mass of the observed molecule. The thermal line width $\Delta {V}_{\rm T} = (8 {\rm ln} 2k{T}_{\rm k}/m)^{1/2}$ for $T_{\rm k}=10$ K is 0.443 km s${}^{-1}$, where $m$ is the mean molecular weight (2.33 amu). Derived line parameters are listed in Table 3. We found that thermal line width for the mean molecular weight is larger than non-thermal line width for most of the observed globules. It is likely that observed globules are dominated by thermal support with smaller contribution from the non-thermal turbulence. The thermal and non-thermal components of the line width can be characterized by a temperature as $T_{\rm k} = \Delta {V}_{\rm T}^{2} m / k 8 {\rm ln} 2$ and $T_{\rm NT} = \Delta {V}_{\rm NT}^{2} m / k 8 {\rm ln} 2$, respectively. We derived $T_{\rm eff} = T_{\rm k}+T_{\rm NT}$ for each globule in Table 3 as a cloud-supporting temperature which includes non-thermal contribution. In order to estimate the distance to the globules, we will use $T_{\rm eff}$ in combination with the Bonnor-Ebert model fitting of globules in \S 4.2.2. In this paper, we only use the information on molecular line width. Detailed studies such as comparison between dust extinction and molecular distribution will be reported in a subsequent paper. 

\section{Density Structure}
\subsection{Bonnor-Ebert Sphere Model}
In order to model the internal density structure of Bok globules, we used the Bonnor-Ebert sphere model (Bonnor 1956; Ebert 1955), which is a modified Lane-Emden equation describing a pressure-confined, self-gravitating isothermal gas sphere in hydrostatic equilibrium. The density structure of a Bonnor-Ebert sphere can be obtained by solving the following equation:
\begin{equation}
{\frac{1}{{ \xi }^{2}}}{\frac{d}{d \xi }}\left({{ \xi }^{2}{\frac{d \phi }{ d \xi }}}\right)={e}^{- \phi }
\end{equation}
where $\phi$ is the logarithmic density contrast, $\phi ( \xi ) = - \ln (\rho / \rho {}_{\rm c})$; $\rho$ and $\rho {}_{\rm c}$ are the volume density and central volume density, respectively. $\xi$ is the dimensionless radial parameter, $\xi = (r/C_{\rm s}) \sqrt {4 \pi G{ \rho }_{\rm c}}$; $r$, $G$, and $C_{\rm s}$ are the radius, gravitational constant, and isothermal sound speed, $C_{\rm s}=(kT/m)^{1/2}$, where $k$, $T$, and $m$ are the Boltzmann constant, kinetic temperature, and mean mass of the molecule (2.33 amu), respectively. Under the standard boundary conditions, $\phi (0)=0$ (i.e., $\rho = \rho {}_{\rm c}$ at $r=0$) and $d\phi (0)/d\xi = 0$ (i.e., $d\rho /dr = 0$ at $r=0$), equation (8) can be solved with numerical integration. If the gas sphere is confined by external pressure $P_{\rm ext}$ at the core boundary radius $R$, the solutions of the Bonnor-Ebert sphere can be characterized by the dimensionless radius $\xi {}_{\rm max} = \xi (r=R)$ as
\begin{equation}
\xi {}_{\rm max} = {{\frac{R}{C_{\rm s}}}}\sqrt {4 \pi G{ \rho }_{\rm c}}.
\end{equation}
$\xi {}_{\rm max}$ is a stability measure of the gas sphere against the gravitational collapse (Bonnor 1956; Ebert 1955). Critical state is achieved at $\xi {}_{\rm max} = 6.5$ corresponding to the density contrast of $\rho {}_{\rm c}/\rho {}_{R} = 14$. For a solution with $\xi {}_{\rm max}> 6.5$, the equilibrium state is unstable to the gravitational collapse. 
The Bonnor-Ebert sphere mass $M$ and the external pressure $P_{\rm ext}$ are calculated as
\begin{eqnarray}
M &=&4 \pi {\alpha }^{ 3}{\rho }_{\rm c}{\xi }_{\rm max}^{ 2}{\left({{\frac{ d \phi }{ d \xi }}}\right)}_{\xi ={ \xi }_{\rm max}} \\
P_{\rm ext} &=& {C_{\rm s}}^{2}{ \rho }_{\rm c}{e}^{- \phi ({ \xi }_{\rm max})}
\end{eqnarray}
where $\alpha = C_{\rm s}/(4 \pi G \rho_{\rm c})^{1/2}$. 

\subsection{Bonnor-Ebert Model Fitting of Bok Globules}
\subsubsection{Derivation of Column Density Profile}
For a comparison of observational data (i.e., $A_{V}$) with a theoretical Bonnor-Ebert density distribution, radial $A_{V}$ profiles for each globule were constructed. We set annuli spaced at 9$''$ around the center of the core that is the $A_{V}$ intensity peak (see Table 4), and calculated averaged $A_{V}$ values at each annulus (see Figure 4). For apparently elliptical globules (CB 87, CB 161, and CB 184) we performed ellipse fitting to the $A_{V}$ distribution, and $\lq \lq$equivalent" radius, $r=\sqrt{ab}$, was substituted for radius where $a$ and $b$ are the respective semi-minor and semi-major axes of the ellipse for the cores. Fitted ellipticity and position angle for the globules are listed in Table 4. 
It is possible that the observed globules with circular projected shapes are actually pole-on prolate ellipsoids. In this case, the shape of the observed $A_{V}$ profile should be steeper than that for the intrinsic shape. Since there is no effective method for estimating how ellongated a globule is along the line of sight, we assumed that the line of sight length of globules is the same as the projected diameter. 
\par
We manually masked extinction features likely to be unrelated to the globules (e.g., the other clouds on the same image, diffuse streaming feature; see Fig. 2), and masked regions were omitted in the circularly averaging procedure. Furthermore we recalibrated the relative zero point of extinction for the analyzed globules by subtracting background extinction (i.e., $A_{V}$ offset value). The $A_{V}$ offsets have been derived from averaging $A_{V}$ within the annulus of 150$''$ to 200$''$ radius, which is sufficiently larger than the primary extinction feature of globules. 
Actually, the subtracted values as $A_{V}$ offset are small (typically $\sim0.5$ mag), and it solely depends on where we set the reference fields in the star count or color excess analysis. 
In order to convert $A_V$ to the H${}_{2}$ column density, we used the relationship between the optical color excess $E_{B-V}$ and the hydrogen column density, 
$N({\rm HI}+{\rm H}_{2})/E_{B-V}=5.8 \times 10^{21}$ ${\rm cm}^{-2}$ ${\rm mag}^{-1}$ (Bohlin, Savage, \& Drake 1978), where $N(H{\rm I}+H_{2})$ = $N({\rm HI})+2N({\rm H}_{2})$. 
This relationship is based on the UV absorption line measurements toward the stars of slightly obscured lines of sight ($E_{B-V} \simlt 0.5$ mag). By assuming that (1) all hydrogen in the cloud is in a molecular form and that (2) the ratio of total-to-selective extinction $R_{V} (\equiv A_{V}/E_{B-V})$ is equal to the standard ISM value of 3.1 (e.g., Whittet 1992), we obtained $N({\rm H}_{2})/A_{V} = 9.4 \times 10^{20}$ ${\rm cm}^{-2}$ ${\rm mag}^{-1}$, and used it to derive the H${}_{2}$ column density. The resulting column density profiles are shown in Figure 4. 

\subsubsection{Derivation of Model Parameters}
In order to produce Bonnor-Ebert model density profiles, we performed numerical integration of equation (8). Since the Bonnor-Ebert sphere is characterized by a single parameter $\xi {}_{\rm max}$ (Eq. [9]), best-fit $\xi {}_{\rm max}$ for each globule can be determined by comparing model column density profiles with those obtained from observations. 
We integrated the model density profile along the lines of sight on grids to calculate the column density profile. In order to match the resolution between the model and the observed column density data, the model column density on the $\Delta \alpha$-$\Delta \delta$ plane was convolved with the beam used in the star count or color excess analysis (cylinder-shaped beam convolved with the Gaussian). In the $\chi {}^{2}$-procedure, the radial profile was fitted from the minimum radius (center) to the radius where the observed $A_{V}$ value reaches $\sim 1$ mag (detection limit). In some cores (Lynds 495 and Lynds 498), deviation of observed profiles from the Bonnor-Ebert sphere is found at diffuse outskirts ($A_{V} \simlt 1$ mag), which was excluded in the fitting. The $\chi {}^{2}$-fitting results are not sensitive to the selection of the maximum radius for fitting part unless there is no significant kink in the column density profile. 
\par
In Figure 4, we show the results of the Bonnor-Ebert fit to the radial column density profile of the globules. The dots and error bars represent the average $N$(H${}_{2}$) values at each annulus of 9$''$ intervals and the rms deviation of data points in each annulus, respectively. The solid line denotes the radial column density profile of the best-fit Bonnor-Ebert sphere which is convolved with beam used in the $A_{V}$ measurements. The dashed line denotes the Bonnor-Ebert model profile before the convolution. In Table 4, we list the derived Bonnor-Ebert parameters for globules including previous reports on four sources (Barnard 68, Barnard 335, Coalsack Globule II, and Lynds 694-2). The error of each parameter in the table represents $1 \sigma$ deviation in the $\chi ^{2}$ density profile fitting. We discuss possible systematic uncertainties in the derivation of $\xi {}_{\rm max}$ in the Appendix. 
\par
Recently, Hotzel et al. (2002b) and Lai et al. (2003) presented detailed discussion on the relation between the physical parameters of the Bonnor-Ebert sphere. Substituting the relations of $R = \theta {}_{R} d \propto d$ and $\rho {}_{\rm c} \propto {N}_{\rm c} ({\rm H}_{2}) \theta {}_{R}^{-1} d^{-1} \propto d^{-1}$ into equation (9), we can obtain $d^{-1}T=$constant (Lai et al. 2003). The distance $d$ or temperature $T$ of globules should be accurately determined from observations. Since there were no accurate methods for measuring distance to the starless, compact, and nearby ($d \simlt$ 500 pc) globules except for the limited cases (e.g., interaction with HII region of known distance, existence of foreground stars toward them), the distances listed in Table 1 serve as the first reliable values. If we use the initially assumed distance $d_{\rm ini}$, the necessary temperature for the Bonnor-Ebert sphere $T_{\rm BE}$ can be determined. If we assume that $T_{\rm eff} (=T_{\rm k}+T_{\rm NT})$ from our molecular line observations (see \S 3.3) can approximate effective supporting temperature of the cloud\footnote{Theoretically, supporting force in a Bonnor-Ebert sphere is purely thermal pressure, but actual clouds also have non-thermal support, e.g., turbulence and magnetic fields, which are widely observed in molecular clouds (see, e.g., Larson 2003). We assume that non-thermal support can be included in the model by replacing the isothermal sound speed $C_{\rm s}=(kT_{\rm k}/m)^{1/2}$ in equation (9) by the effective sound speed $C_{\rm eff}=(kT_{\rm eff}/m)^{1/2}$.}, the corrected distance to the globule, $d = (T_{\rm eff}/T_{\rm BE})d_{\rm ini}$, can be determined. 
Finally we derived all the physical parameters of best-fit Bonnor-Ebert spheres (e.g., radius, central density, and mass) for the above $\lq \lq$distance-assumed" and $\lq \lq$temperature-assumed" cases as listed in Table 5. We note that the $\lq \lq$temperature-assumed" distance to the starless globule FeSt 1-457 is estimated to be $\sim$ 70 pc, which would make this globule one of the nearest dark clouds. 

\section{Discussion: Stability and Evolution of Globules}
In this section, we discuss the internal structure and stability of globules by using the data of the ten globules from our observations and four from the literature. 

\subsection{Physical Properties of Globules}
Though we have estimated physical properties of globules by assuming the Bonnor-Ebert model, there are actually other theoretical models (e.g., Shu 1977; Larson 1969; Penston 1969; McLaughlin \& Pudritz 1997; Plummer 1911). Bacmann et al. (2000) presented that a Bonnor-Ebert-like model (a finite size sphere with inner uniform density region and $\rho \propto r{}^{-2}$ outer envelope) fits well with the column density profile of globules obtained from mid-infrared 7 $\mu$m absorption measurements, but that a singular-isothermal-sphere (SIS) or singular logotrope sphere cannot reproduce the observed profile well. By considering the resolution of our extinction maps ($\sim$30$''$) and possible geometric error of globules (e.g., deviation from spherical symmetry), it is not possible to make meaningful comparisons between the Bonnor-Ebert model and other Bonnor-Ebert-like models (see, e.g., Figure 2 in Harvey et al. 2003, which compares the Bonnor-Ebert model with a Plummer-like model). Here, we simply assume the Bonnor-Ebert model. 
\par
As described in the previous section, the dimensionless radial parameter $\xi {}_{\rm max}$ determines the shape of a Bonnor-Ebert density profile as well as the stability of the equilibrium state against the gravitational collapse. 
We plot globules on the $\xi {}_{\rm max}$ versus density contrast diagram in the upper panel of Figure 5. Since $\xi {}_{\rm max}$ has one-to-one correspondence with density contrast, all the Bonnor-Ebert spheres are distributed along the solid curved line. We found that more than half of starless globules (7 out of 11 sources) are located near the critical state,  $\xi {}_{\rm max}=6.5 \pm 2$. Thus, we suggest that a nearly critical Bonnor-Ebert sphere characterizes the typical density structure of starless globules. The remaining starless globules show clearly unstable states ($\xi {}_{{\rm max}} > 10$). When we divide starless globules into two groups with respect to the critical line ($\xi {}_{\rm max} = 6.5$), there are three stable starless globules and eight unstable starless globules. The majority of starless globules is located in the unstable states ($\xi {}_{\rm max} > 6.5$) if the uncertainties in $\xi {}_{\rm max}$ are not taken into account. We also found that all the star-forming globules have larger $\xi {}_{\rm max}$ values ($>$ 10) than that of the critical Bonnor-Ebert equilibrium state, which is consistent with the fact that they have started gravitational collapse. In the lower panel of Figure 5, we show the fitting error in $\xi {}_{\rm max}$ for each globule. Filled gray circles denote the fitting results for some globules with considerable features of ambient extinction if the masked region on the $A_V$ map (see Fig 2) is included in the derivation of the column density profile. In addition to the $\xi {}_{\rm max}$ diagram in Figure 5, we made histograms of logarithmic density contrast for starless and star-forming globules (Figure 6). Similar characteristics as descrived above can also be seen in the histograms. 
\par
To see the relationships between derived physical properties of globules in Tables 4 and 5, we made correlation diagrams between density contrast and the other physical parameters of globules in Figure 7. We consider that the horizontal axis (density contrast) represents the evolutionary states of globules, because globules should evolve toward higher central condensation to form stars. We can see from panels (e) and (f) in the figure that temperature $T_{\rm eff}$ and external pressure $P_{\rm ext}$ of globules appear roughly constant regardless of the density contrast. 
If we assume constant $T_{\rm eff}$ and $P_{\rm ext}$, the Bonnor-Ebert model parameters for each density contrast value can be constrained. The broken line in each panel is the relationship for the Bonnor-Ebert spheres with a constant $T_{\rm eff}$ of $13.2$ K and $P_{\rm ext}$ of $5.3 \times 10^{4}$ K cm${}^{-3}$, which are the median values for the globule samples. The value of $P_{\rm ext}$ is larger than the previously reported ISM pressure in the solar vicinity of $\sim 1.8 \times 10^{4}$ K cm${}^{-3}$ (McKee 1999). 
\par
The overall trend in the data point distributions in Figure 7 can be approximated by Bonnor-Ebert spheres with a constant $\lq \lq$effective'' temperature and a constant external pressure. The relations for the radius and mass (panels [c] and [d]) remain roughly constant against the density contrast, suggesting that these quantities are not sensitive to the evolutionary states of globules. This result agrees with the N${}_{2}$H${}^{+}$ observations of molecular cloud cores in Taurus (Tatematsu et al. 2004), in which there is no remarkable difference in core radius and mass between starless cores and cores with stars. Since the density at core boundary is derived from $P_{\rm ext} / T_{\rm eff}$, the density at core center $n_{\rm c}$ is directly obtained for each density contrast value and shows a proportional relationship against the density contrast as shown in panel (b). Accordingly, the density contrast and $A_{V}$ show positive correlation. The data points in panel (a) are located slightly lower than the broken line because of a beam dilution effect in the $A_{V}$ measurements with $\sim 30''$ resolution. For molecular cloud cores located in regions of higher pressure (e.g., dark cloud complex) than in the solar vicinity, the Bonnor-Ebert relationship in panels (a) and (b) should be shifted upward corresponding to the larger value of $P_{\rm ext} / T_{\rm eff}$. We can see from panel (a) that the density contrast value is simply estimated from the $A_{V}$ (column density) measurements under the assumption of the Bonnor-Ebert sphere with a constant $T_{\rm eff}$ and $P_{\rm ext}$. 

\subsection{On the Selection Bias of the Globule Samples}
Here we briefly consider the selection bias in our globule samples. The peak $A_{V}$ values for our observed globules range from $\sim 6.5$ to $41$ mag. We selected globules that were opaque at optical wavelengths, i.e., with few stars in the background of the cloud on the Digitized Sky Survey image. These sources are expected to have $A_{V}$ greater than at least 5 mag, and expected to be sites of future star formation, i.e., so-called $\lq \lq$dense cores''. The lower limit of the peak $A_{V}$ for the globules may be determined by the criterion in our target selection. Thus, it is possible that the diffuse cores of low column densities are not included in our globule sample as in the case of previous studies on dense cores (e.g., Visser et al. 2002; Lee \& Myers 1999; Jijina et al. 1999). We note that the lower limit $A_{V}$ of  $\sim 5$ mag corresponds to a low density contrast of $\sim 3$ if we assume a constant $T_{\rm eff}$ and $P_{\rm ext}$ used in the previous section. The density contrast value is much less than the critical value of $14$.

\subsection{Implications for the Stability of Globules}
It is interesting that there are starless globules showing larger density contrast than that of the maximum value ($\sim$14) for stable Bonnor-Ebert spheres (see $\S$ 5.1), because unstable equilibrium states should not last long without any extra stabilizing force such as magnetic pressure and/or turbulent pressure (e.g., McKee \& Holliman 1999). Since unstable equilibrium states cannot last long, the globules of large density contrast should not exactly be the Bonnor-Ebert sphere even if they mimic the Bonnor-Ebert density profile. Similar results have also been reported from the mid-infrared extinction study of starless dense cores (density contrasts are generally $\sim$10$-$80; see Table 3 of Bacmann et al. 2000). If we consider a simple case in which the non-thermal pressure components have the same radial dependence as the thermal pressure, the theoretical Bonnor-Ebert model can be modified by replacing its kinetic temperature term with $T_{\rm eff}$ as described in $\S$ 4.2.2. 
In this case, the shape of the Bonnor-Ebert density profile as well as the stability criteria (i.e., the value of $\xi {}_{\rm max}$ or maximum density contrast) hold unchanged compared with the purely thermally supported case. This modification cannot provide a realistic explanation for the stability of the large density contrast globules, suggesting either that they are already collapsing toward higher central condensation as discussed in $\S$ 5.5, or that the stabilizing mechanisms accounting for their large density contrast are more complex. The Bonnor-Ebert assumption for the globules is not valid in these cases. We note that our result is in marked contrast to one in which most of the clumps observed at 850 $\mu$m dust emission in the Orion B molecular cloud complex are in stable equilibrium based on the Bonnor-Ebert analysis (Johnstone et al. 2001). 
\par
A non-isentropic multi-pressure polytrope model (McKee \& Holliman 1999) may provide successful interpretation allowing larger density contrast than that of the critical Bonnor-Ebert sphere. Curry \& McKee (2000) developed a composite polytrope model which also allows large density contrast. These two models provide a stable equilibrium solution of arbitrarily large density contrast by adjusting the contributions from the non-thermal pressure components, i.e., turbulent and magnetic pressure. Galli et al. (2002) studied the effects of the interstellar radiation field (ISRF) on the cloud stability. They reported that a cloud's maximum density contrast increases with increasing intensity of the ISRF if the stabilizing effect of external heating of the cloud as well as produced temperature gradient are considered. These models are possible alternatives to the Bonnor-Ebert sphere providing larger value of maximum density contrast ($>14$). The shape of their density profiles have characteristics similar to those of the Bonnor-Ebert sphere (see, Fig. 3 and 4 in Galli et al. 2002, Fig. 10 in McKee \& Holliman 1999), that is a flat inner region surrounded by steeper envelope decreasing with power-law, in particular $\rho \propto r^{- \alpha}$ where $\alpha \sim 2$.  It seems difficult to distinguish them only from the observed shape of the density profile. In other words, each model should provide similar density profiles and corresponding density contrast values for the same observational dataset. Detailed observations of velocity and temperature structure of dense cores are needed to determine the most realistic model. 
\par
Major differences among the models discussed above are the supporting mechanisms of the cloud and the stability criterion, i.e. the value of the critical density contrast. The actual density contrast value for globules can be obtained through the Bonnor-Ebert fit (see Table 4), but its critical value is not always known (i.e., model dependent). Initial density structure of globules right before the collapse can be determined if we constrain the value of its critical density contrast. Initial density conditions for star formation will provide a good starting point for the theoretical calculations. For example, given initial density profile of a globule, time evolution of mass accretion rate can be calculated (e.g. Whitworth \& Ward-Thompson, 2001), which is an important parameter for constraining the mass of a newly formed star. 
\par
In order to estimate the value of critical density contrast from observations, it should be worth investigating the frequency distribution of density contrast for a number of starless globules. Since unstable equilibrium states should not be long sustainable, we expect that the detection probability for stable globules is much higher than that for unstable ones. Thus, it is likely that the frequency distribution of density contrast shows a sharp decrease toward larger density contrasts at a certain bin, which should correspond to the density contrast critical for collapse. Though the number of our samples (11 sources) is not large, we present the histogram of density contrast for the starless globules in Figure 6. The density contrast of the starless globules is mostly populated near the critical Bonnor-Ebert sphere and decreases toward the larger density contrast, suggesting that the actual critical density contrast of globules is not very different from that for the Bonnor-Ebert sphere. A statistical study with a larger number of samples is our future plan. 

\subsection{Starless Globules with Stable Bonnor-Ebert Solutions}
In Table 4, three out of fourteen globules are fitted as a stable equilibrium state of the Bonnor-Ebert sphere. If they are truly stable, how can stable globules form stars? Stable globules might increase their density contrast quasi-statically toward the onset of dynamical collapse with gradually increasing external pressure (e.g., Hennebelle et al. 2003) and/or with decreasing internal turbulent pressure (Nakano 1998). For the case of rapid increase of external pressure, core collapse triggered by a compression wave should be taken into account (e.g., Hennebelle et al. 2003, 2004; Motoyama et al. 2004). Our observations of globules, however, show neither systematic decrease in $T_{\rm eff}$ nor increase in $P_{\rm ext}$ (panels e and f of Fig. 7) toward larger density contrast. It seems that non-thermal turbulence is already mostly dissipated even in the stable globules. Though there is only small room for the dissipation of turbulence, a slight decrease of turbulence may initiate the collapse of stable globules as suggested by Nakano (1998). The decrease of $T_{\rm eff}$ (i.e., line widths) necessary for stable globules to start gravitational collapse is small. This is consistent with the result that there is no apparent tendensity of systematic decrease in $T_{\rm eff}$ against density contrast (panel e of Fig. 7). In the star forming regions where non-thermal motions are more prominent than globules, e.g., GMCs, the dissipation of turbulence in dense cores is reported through the virial analysis (e.g., OMC-2/3 region in Orion: Aso et al. 2000). 
\par
Since possible evolutionary processes of stable globules are expected not to be dynamical, their lifetime should be long compared with unstable ones. However, in our starless globule samples, the fraction of stable globules (density contrast is less than 14) is small (3 out of 11 sources). The observed starless globules are mostly populated near the critical Bonnor-Ebert sphere, and decreases toward the smaller density contrast (Figure 6). This result is inconsistent with an idea that all of nearly critical globules originated from stable low-density-contrast globules; if the idea is the case, the number of the observed low-density-contrast globules should be larger than or comparable with that of the nearly critical ones. Thus, it is most likely that most of the nearly critical globules which we observed are not evolved from stable globules. The origin of nearly critical globules can be related to the formation mechanism of globules. Further studies on this subject are desirable. 

\subsection{Slow Collapse of A Nearly Critical Bonnor-Ebert Sphere} 
After gravitational collapse begins, globules evolve toward higher central condensation. If we assume an extreme case in which all the globules fitted as unstable Bonnor-Ebert equilibrium states are already collapsing, the difference in $\xi {}_{\rm max}$ is attributed to different evolutionary stages of globules. Since the Bonnor-Ebert spheres provide a good fit to starless globules, it is meaningful to consider the gravitational collapse of the Bonnor-Ebert sphere (e.g., Foster \& Chevalier 1993; Ogino, Tomisaka, \& Nakamura 1999). We expect that the critical Bonnor-Ebert sphere can approximate the initial condition of collapse, because a large fraction of the starless globules are located near the critical state, $\xi {}_{\rm max}=6.5 \pm 2$, as shown in $\S$ 5.1. 
\par
In order to make quantitative comparisons of the density structures between our observations and a collapsing gas sphere starting from a nearly critical Bonnor-Ebert sphere, we used the result of theoretical calculations from Aikawa et al. (2005). 
In their calculation, the central density $n_{\rm c} (\rm{H})$ and kinetic temperature $T$ of the critical Bonnor-Ebert sphere were set as $n_{\rm c} (\rm{H})=2 \times 10^4$ $\rm{cm} {}^{-3}$ and $T=10$ K, respectively. No cloud rotation, magnetic fields, or non-thermal turbulence were considered, and a fixed boundary condition of zero velocity was adopted at the outermost radius of $R = 0.2$ pc, which is slightly larger than the critical radius of an equilibrium sphere (see, Aikawa et al. 2005). We note that the initial condition parameters of $n_{\rm c} (\rm{H})$, $T$, and $R$, are consistent with the observed physical properties of globules around $\xi {}_{\rm max}=6.5$ (see, Table 4 and 5). 
The initial density distribution was multiplied by a constant factor $\alpha$ of 1.1 in order to initiate collapse. The slight density enhancement of $\alpha = 1.1$ is preferable to approximate the collapse from a nearly critical Bonnor-Ebert sphere. The time scale of the collapse is several times the free-fall time, $\sim 1.2 \times 10^{6}$ yr ($\sim 4 \times t_{\rm ff}$), which is comparable with the observational lifetime of starless dense cores ($\sim 10^{6}$ yr; e.g., Visser et al. 2002; Lee \& Myers 1999). 
\par
Figure 8 shows the evolution of a radial column density profile of the collapsing sphere for the case of $\alpha =1.1$ described above. The gray solid line denotes initial column density profile ($t = 0$ yr), and the black solid, dotted, and broken lines denote density profile at specific times of $4.4 \times 10^5$, $8.9 \times 10^5$, and $1.1 \times 10^6$ yr, respectively. We found that the column density profiles of the collapsing sphere are well fitted with the static Bonnor-Ebert spheres (equilibrium solutions) in the range of $\xi {}_{\rm max} < 25$ within the relative difference of 2\%. The plus symbols show the best-fitting result for each collapsing profile with the Bonnor-Ebert sphere, which are labeled with corresponding $\xi {}_{\rm max}$ values; for example, the density profile of the collapsing sphere at $4.4 \times 10^5$ yr mimics the unstable Bonnor-Ebert sphere of $\xi {}_{\rm max} = 7.6$. This result suggests that observed starless globules with large density contrasts are gravitationally collapsing objects. Figure 9 shows the relationship between $\xi {}_{\rm max}$ versus density contrast. The solid line is for the Bonnor-Ebert sphere, and the dotted line is for the collapsing sphere which plots the best-fitting $\xi {}_{\rm max}$ values against the density contrast of the collapsing sphere. The plus symbols denote the elapsed time of the collapsing sphere at 10\% intervals of the total collapse time. 
Since the free-fall time scale increases with decreasing density, a collapsing gas sphere starting from a nearly critical Bonnor-Ebert sphere should mimic marginally unstable Bonnor-Ebert spheres for a long time. Thus, this is qualitatively consistent with the fact that the $\xi {}_{\rm max}$ for the starless globules is mostly populated near the critical state (7 out of 11 sources) and the remaining starless globules show clearly unstable states ($\xi {}_{\rm max} > 10$; 4 out of 11 sources) as shown in $\S$ 5.1. From the model prediction, we note that 50\% of collapsing globules should be located in $\xi {}_{\rm max} = 6.5 - 8.5$, and the rest of the globules should have $\xi {}_{\rm max}$ larger than 8.5. The fraction is consistent with our result that half of the starless globules with $\xi {}_{\rm max} > 6.5$ are located in $\xi {}_{\rm max} = 6.5 - 8.5$ (4 out of 8 sources, see Fig. 5). In Figure 10, we demonstrate the histogram of logarithmic density contrast for starless globules (the same as Figure 6) and the time evolution of the model collapsing sphere (solid line with dots). Each dot represents the residence time of a collapsing sphere measured at 0.2 intervals in the logarithmic density contrast. The histogram of the observed globule in the unstable region is consistent with that from model calculations of a collapsing sphere. 
\par
As discussed above, the density structure of the observed globules can be explained in terms of the slow collapse of a nearly critical Bonnor-Ebert sphere. The anticipated inward velocity from the model calculation remains subsonic\footnote{In $\S$ 4.2.2, we have estimated distance to the globules using $T_{\rm eff}$ from radio molecular line observations as shown in Table 3. A collapsing motion in the globules should contribute to the derivation of $T_{\rm eff}$ if the motion is large enough compared with the observed molecular line width. Since predicted inward velocity from the model is small, it is likely that the contribution of collapsing motion to the derived distance is negligible.} 
for a long time; the peak inward velocity is less than $\sim 0.1$ km s${}^{-1}$ for $\sim 10^6$ yr, which is $\sim 80$\% of the total collapse time (see, Fig. 1 of Aikawa et al. 2005). Previous reports on the observations of CS ($J=2 \rightarrow 1$) (blue) asymmetry profile, indicative of gravitational collapse in dense cores, show that infall candidates constitute 25 \% of the starless core samples (Lee, Myers, \& Tafalla 1999). It is possible that the rest of the starless cores have a slow collapse motion which is hard to detect. Radio molecular line observations of infall asymmetry profiles with sufficiently high velocity resolution (e.g. $\sim 0.01$ km s${}^{-1}$) are necessary to confirm the scenario of slowly collapsing globules. Among the starless globules in Figure 5, the gas inward velocity for Lynds 694-2 ($\xi {}_{\rm max} = 25 \pm 3$) was measured as $0.05 - 0.07 $ km s${}^{-1}$ (Lee, Myers, \& Tafalla 2001; Lee, Myers \& Plume 2004), which is consistent with the prediction from the collapse model ($\alpha = 1.1$). 

\section{Summary}
We have carried out a near-infrared ($J$, $H$, and $K_{s}$) imaging survey of ten Bok globules using the infrared camera SIRIUS on the IRSF 1.4 m telescope in South Africa. $A_{V}$ distributions for the ten globules were derived through measurements of the $H-K$ color excess or the stellar density at $J$, $H$, and $K_{s}$ (star count). 
The radial column density profile for each globule was analyzed using the Bonnor-Ebert sphere model, which well fits the profile within observational uncertainties, and physical properties of globules, e.g., size, central density, temperature, mass, external pressure, and center-to-edge density contrast, were derived. The dimensionless radial parameter $\xi {}_{\rm max}$ determines the shape of a Bonnor-Ebert density profile as well as the stability of the equilibrium state, e.g., the solution of $\xi {}_{\rm max} > 6.5$ is unstable to the gravitational collapse. We investigated the stability of globules on the basis of $\xi {}_{\rm max}$ for ten globules from our observations and four globules in the literature. In addition to the near-infrared imaging, we have carried out radio molecular line observations toward the same ten Bok globules using the 25-element focal-plane SIS receiver BEARS on the 45 m telescope of Nobeyama Radio Observatory (NRO). We measured line width of the globules for independent measurements of the $\lq \lq$effective" temperature including the contribution of turbulence. It was confirmed that most of the globules are dominated by thermal support. Since the distance-dependent parameters of the Bonnor-Ebert sphere were derived using the initially assumed distance to the globules, derived best-fitting temperature was affected by the distance assumption. We constrained the distance to the globules by comparing the temperature from the molecular line width with the temperature from the Bonnor-Ebert fitting, and rescaled distance-dependent physical quantities of globules. 
\par
(1) We found that more than half of the starless globules (7 out of 11 sources) are located near the critical state, $\xi {}_{\rm max}=6.5 \pm 2$. Thus, we suggest that a nearly critical Bonnor-Ebert sphere characterizes the typical density structure of starless globules, and it approximates the initial condition of gravitational collapse. 
\par
(2) We found that four out of eleven starless globules show clearly unstable states ($\xi {}_{{\rm max}} > 10$). 
Since unstable equilibrium states should not be long sustained, we expect that they are already collapsing toward higher central condensation or that extra force (e.g., magnetic and/or turbulent pressure) accounting for large $\xi {}_{\rm max}$ stabilizes the globules. It was also found that all three star-forming globules have unstable solutions of $\xi {}_{\rm max} > 10$, which is consistent with the fact that they have started gravitational collapse. 
\par
(3) We investigated the collapse of the Bonnor-Ebert sphere from a nearly critical state using the model calculation of Aikawa et al. (2005), and found that the column density profiles of the collapsing sphere mimic those of static Bonnor-Ebert spheres (unstable equilibrium solutions). By relating $\xi {}_{\rm max}$ to the collapsing sphere at a specific time, the evolutionary state of globules can be interpreted, and the detection probability of each $\xi {}_{\rm max}$ value can be predicted from the model calculation. Since the evolutionary timescale decreases with increasing density, the collapsing sphere resembles a marginally unstable Bonnor-Ebert sphere for a long time. It was found that the frequency distribution of $\xi {}_{\rm max}$ for the observed starless globules is consistent with that from model calculations of the collapsing sphere. 

\acknowledgements
\paragraph{Acknowledgements \\}
We are grateful to Masao Saito and Koji Tomisaka for their helpful comments and suggestions. Thanks are due to the staff at SAAO and NRO for their kind support during the observations. We also thank Doug Johnstone, the referee of this paper, for useful comments on the manuscript. 
The IRSF/SIRIUS project was initiated and supported by Nagoya University, National Astronomical Observatory of Japan, and University of Tokyo in collaboration with South African Astronomical Observatory under a financial support of Grant-in-Aid for Scientific Research on Priority Area (A) No. 10147207 and No. 10147214, and Grant-in-Aid No. 13573001 of the Ministry of Education, Culture, Sports, Science, and Technology of Japan. M. T. acknowledges support by the Grant-in-Aid (No. 12309010, 16340061, 16077204). Y. A. is supported by Grant-in-Aid for Scientific Research (No. 14740130, 16036205) and $\lq \lq$The 21st Century COE Program of Origin and Evolution of Planetary Systems" of the Ministry of Education, Culture, Sports, Science and Technology of Japan (MEXT). 

\appendix
\paragraph{\bf Appendix A: Possible Systematic Uncertainties in $\xi {}_{\rm max}$ \\}
Here we discuss the ambiguity of derived column density ($A_{V}$), which leads to the systematic uncertainties in $\xi {}_{\rm max}$. In order to evaluate the dependence of the conversion factor, $A_{V} / E_{H-K}$ and $A_{V} / A_{\lambda}$ $(\lambda = J$, $H$, and $K)$, on the ratio of total-to-selective extinction $R_{V}$, we used the $R_{V}$-dependent extinction law empirically derived by Cardelli et al. (1989). Though it is well established that $R_{V}$ has a uniform value of $\sim$3.1 in the ISM (e.g., Whittet 1992), there are some observational reports on higher $R_{V}$ values up to $\sim 6$ in dense molecular clouds (e.g., Kandori et al. 2003, and references therein). The empirical extinction law for $R_{V}=6$ gives the conversion factors smaller $\sim$ 20 \% than those for the case of $R_{V}=3.1$, which we adopted in the analysis (\S 3.1 and \S 3.2). Note that the extinction law gives slightly different conversion factors compared with Rieke \& Lebofsky (1985) as shown in Table 3 of Cardelli et al. (1989). This possible systematic uncertainty as well as the uncertainty of the gas-to-dust ratio can be included in the derivation of $N({\rm H}_{2})$. We have converted $A_{V}$ to H${}_{2}$ column density using the relationship $N({\rm H}_{2}) / A_{V} = 9.4 \times 10^{20}$ cm${}^{-2}$ mag${}^{-1}$ derived from Bohlin, Savage, \& Drake (1978). Similar relationships were obtained from the X-ray absorption studies toward SNRs with high dynamic range ($A_{V} \simlt 30$ mag) as 
$N_{\rm H}/A_{V}=1.8-2.2 \times 10^{21} {\rm cm}^{-2} {\rm mag}^{-1}$ 
(Gorenstein 1975; Ryter et al. 1975; Ryter 1996; Predehl \& Schmitt 1995). 
Thus, the conversion coefficient is likely to be determined within $\sim 20$ \% error. 
It can be seen from the panel (a) of Figure 7 that $A_{V}$ is roughly correlated with about 1/2th power of density contrast for the Bonnor-Ebert spheres. Thus, for the case that $A_{V}$ is systematically changed by a factor $\beta$, the value of density contrast varies by $\beta {}^{1.5}$ accordingly. Since $\xi {}_{\rm max}$ is proportional to $\sqrt{\rho {}_{\rm c}}$, corresponding $\xi {}_{\rm max}$ value is expected to vary by $\beta {}^{0.75}$. 
If we assume that derived $A_{V}$ for each globule is overestimated by 20 \%, $\xi {}_{\rm max}$ is overestimated by $\sim$ 15 \%. This is comparable to the uncertainty in $\xi {}_{\rm max}$ derived from the $\chi {}^{2}$ fitting, and this modification should not significantly change our conclusions on the density structure of globules.

\begin{landscape}
\begin{table}
\tablewidth{500pt}
\begin{center}
\caption{Source list}
\begin{tabular}{lccclll}
\tableline\tableline
Name & 
R.A. (J2000)\tablenotemark{a} & 
Dec. (J2000)\tablenotemark{a} & 
Distance & 
IRAS\tablenotemark{b} & 
Other name & 
Reference\tablenotemark{c} 
\\
  & 
($^{\rm h}$ $^{\rm m}$ $^{\rm s}$) & 
($^{\circ}$ $^{'}$ $^{''}$) & 
(pc) & 
  & 
  & 
\\
\tableline
CB 87   & 17 25 05 & $-$24 07 19 & 160 & No & Barnard 74, Lynds 81 & 1,2 \\
CB 110 & 18 05 55 & $-$18 25 10 & 180 & No & Lynds 307 & 1,3,4 \\
CB 131 & 19 17 00 & $-$18 01 52 & 180 & No & Barnard 93, Lynds 328 & 1,3,4 \\
CB 134 & 18 22 45 & $-$01 42 40 & 260 & No & & 1,5,A \\
CB 161 & 18 53 56 & $-$07 26 29 & 400 & No & Barnard 118, Lynds 509 & 1,6 \\
CB 184 & 19 31 52 & $+$16 27 14 & 300 & No\tablenotemark{d} & Lynds 709 & 1,3,4 \\
CB 188 & 19 20 16 & $+$11 36 15 & 300 & Yes\tablenotemark{e} & Lynds 673-1 & 1,3,4 \\
FeSt 1-457 & 17 35 45 & $-$25 33 11 &160 & No & & 1,3,4,B \\
Lynds 495 & 18 38 58 & $-$06 44 00 & 200 & No & & 1,7,C \\
Lynds 498 & 18 40 11 & $-$06 40 45 & 200 & No & & 1,7,C \\
\tableline
\end{tabular}
\tablenotetext{a}{Position of field center for each near-infrared image.}
\tablenotetext{b}{Existence of IRAS point sources within optical boundary of each core.}
\tablenotetext{c}{(1) Dutra \& Bica (2002); (2) Huard, Sandell \& Weintraub (1999); (3) Launhardt \& Henning (1997); (4) Dame et al. (1987); (5) Straizys, Cernis, \& Bartasiute (1996); (6) Leung, Kutner, \& Mead (1982); (7) Schneider, \& Elmegreen (1979); (A) Assumed to have the same distance as Serpens molecular cloud; (B) Assumed to have the same distance as Barnard 83; (C) Assumed to have the same distance as GF 5 dark cloud filament. }
\tablenotetext{d}{IRAS 19116+1623 (Class II located near core boundary): Launhardt (1996)}
\tablenotetext{e}{IRAS 19179+1129 (Class I with outflow): Launhardt (1996), Yun, \& Clemens (1992)}
\end{center}
\end{table}
\end{landscape}

\begin{landscape}
\begin{table}
\begin{center}
\caption{Results of stellar detection}
\begin{tabular}{llccclccclcccc}
\tableline\tableline
Field & & \multicolumn{3}{c}{Limiting mag.\tablenotemark{a}} & &  \multicolumn{3}{c}{Threshold mag.\tablenotemark{b}} & & \multicolumn{4}{c}{Detected number of stars\tablenotemark{c}}  \\
\cline{3-5} \cline{7-9} \cline{11-14}
         & & $J$ & $H$ &  $K_{s}$ & & $J$ & $H$ &  $K_{s}$ & & 
\multicolumn{1}{c}{$J$} & 
\multicolumn{1}{c}{$H$} &  
\multicolumn{1}{c}{$K_{s}$} & 
\multicolumn{1}{c}{$H$ and  $K_{s}$}   
\\
\tableline
CB 87   & & 18.9 & 18.4 & 17.5 & & 18.5 & 17.8 & 17.1 & & 8107 & 10588 & 6510 & 7609  \\
CB 110 & & 18.8 & 17.9 & 17.1 & & 18.1 & 17.2 & 16.4 & & 12471 & 14220 & 12340 & 12119  \\
CB 131 & & 19.1 & 18.6 & 17.5 & & 18.1 & 17.4 & 16.8 & & 6678 & 10760 & 111868 & 9890  \\
CB 134 & & 19.5 & 19.2 & 17.9 & & 18.9 & 18.5 & 17.5 & & 5631 & 7530 & 4530 & 5510  \\
CB 161 & & 18.5 & 18.3 & 17.3 & & 18.0 & 17.7 & 16.9 & & 6431 & 8406 & 5330 & 6044  \\
CB 184 & & 18.8 & 18.1 & 17.3 & & 18.3 & 17.8 & 17.1 & & 5235 & 6056 & 4241 & 4673  \\
CB 188 & & 18.9 & 18.3 & 17.4 & & 18.4 & 17.6 & 17.0 & & 6743 & 8963 & 7372 & 7669  \\
FeSt 1-457 & & 19.0 & 18.2 & 17.4 & & 18.6 & 17.2 & 16.8 & & 9261 & 10614 & 10444 & 9240  \\
Lynds 495 & & 19.1 & 18.0 & 16.6 & & 18.4 & 17.2 & 15.9 & & 10764 & 13404 & 10510 & 10760  \\
Lynds 498 & & 18.7 & 17.6 & 16.7 & & 17.7 & 16.8 & 16.0 & & 13826 & 13696 & 12064 & 11673  \\
\tableline
\end{tabular}
\tablenotetext{a}{Limiting magnitude is defined by the magnitudes whose photometric error is equal to 0.1 mag.}
\tablenotetext{b}{Threshold magnitude is defined by the peak value of the histogram distribution of magnitudes for detected stars whose magnitude errors are less than 0.1 mag.}
\tablenotetext{c}{The first three columns show the number of stars detected on the $J$, $H$, or  $K_{s}$ images (single band detection) whose photometric error is less than threshold magnitudes. 
The last column shows the number of stars detected both on the $H$ and  $K_{s}$ images whose photometric errors, $\sqrt {\delta H^{2}+\delta K_S^{2}}$, are less than 0.15 mag.} 
\end{center}
\end{table}
\end{landscape}

\clearpage
\begin{landscape}
\begin{table}
\tablewidth{600pt}
\begin{center}
\caption{Molecular line parameters of globules }
\begin{tabular}{lllllclcc}
\tableline \tableline
Name &	
\multicolumn{1}{c}{$V_{\rm LSR}$} &	
\multicolumn{1}{c}{$\Delta V$\tablenotemark{a}} &	
\multicolumn{1}{c}{$\Delta V_{\rm NT}$\tablenotemark{b}} & 	
$T_{\rm eff}$\tablenotemark{c} & 	
\multicolumn{1}{c}{$\tau$\tablenotemark{d}} & 	
$T_{\rm ex}$\tablenotemark{e} & 	
Molecular Line  
\\
  &	
\multicolumn{1}{c}{(km s${}^{-1}$)} &	
\multicolumn{1}{c}{(km s${}^{-1}$)} &	
\multicolumn{1}{c}{(km s${}^{-1}$)} & 	
(K) & 	
   & 	
(K) & 	
\\
\tableline
CB 87\tablenotemark{f}   
                  & 4.73$\pm$0.02 & 0.21$\pm$0.04 & 0.17$\pm$0.05 & 11.4$\pm$0.8 &
                  0.55$\pm$0.12 & 10 & C$^{18}$O ($J=1-0$) \\
				  & 4.73$\pm$0.01 & 0.25$\pm$0.02 & 0.22$\pm$0.03 & 12.4$\pm$0.6 & 
				  0.38$\pm$0.04 & 10 &   \\
CB 110      & 5.94$\pm$0.02 & 0.50$\pm$0.04 & 0.48$\pm$0.04 & 21.8$\pm$2.2 &
                  0.62$\pm$0.06 & 10 & C$^{18}$O ($J=1-0$) \\
				  & 5.91$\pm$0.01 & 0.45$\pm$0.02 & 0.43$\pm$0.02 & 19.3$\pm$1.0 & 
				  0.47$\pm$0.02 & 10 &   \\
CB 131      & 6.66$\pm$0.01 & 0.56$\pm$0.02 & 0.55$\pm$0.02 & 25.1$\pm$1.3 & 
                  0.58$\pm$0.03 & 10 & C$^{18}$O ($J=1-0$) \\
				  & 6.65$\pm$0.01 & 0.57$\pm$0.02 & 0.56$\pm$0.02 & 25.8$\pm$1.1 & 
				  0.56$\pm$0.02 & 10 &   \\
CB 134      & 8.56$\pm$0.01 & 0.28$\pm$0.02 & 0.25$\pm$0.02 & 13.2$\pm$0.6 & 
                  0.94$\pm$0.10 & 10 & C$^{18}$O ($J=1-0$) \\
				  & 8.562$\pm$0.004 & 0.24$\pm$0.01 & 0.20$\pm$0.01 & 12.1$\pm$0.2 & 
				  0.81$\pm$0.04 & 10 &   \\
CB 161      & 12.48$\pm$0.01 & 0.25$\pm$0.02 & 0.22$\pm$0.02 & 12.5$\pm$0.6 & 
                  0.61$\pm$0.06 & 10 & C$^{18}$O ($J=1-0$) \\
				  & 12.47$\pm$0.01 & 0.27$\pm$0.01 & 0.24$\pm$0.02 & 13.0$\pm$0.4 & 
				  0.44$\pm$0.02 & 10 &   \\
CB 184      & 6.10$\pm$0.01 & 0.35$\pm$0.02 & 0.33$\pm$0.02 & 15.5$\pm$0.6 & 
                  1.01$\pm$0.08 & 10 & C$^{18}$O ($J=1-0$) \\
				  & 6.096$\pm$0.005 & 0.35$\pm$0.01 & 0.33$\pm$0.01 & 15.5$\pm$0.4 & 
				  0.88$\pm$0.04 & 10 &   \\
CB 188      & 6.98$\pm$0.01 & 0.44$\pm$0.03 & 0.42$\pm$0.03 & 19.0$\pm$1.3 & 
                  0.54$\pm$0.04 & 10 & C$^{18}$O ($J=1-0$) \\
				  & 6.92$\pm$0.01 & 0.52$\pm$0.02 & 0.50$\pm$0.02 & 22.9$\pm$1.2 & 
				  0.37$\pm$0.02 & 10 &   \\
FeSt 1-457  & 5.820$\pm$0.003 & 0.182$\pm$0.006 & 0.132$\pm$0.009 & 10.9$\pm$0.1 & 
                  9.56$\pm$1.29 & 6.25$\pm$0.21 & N$_{2}$H$^{+}$ ($J=1-0$) \\
				  & 5.824$\pm$0.002 & 0.175$\pm$0.004 & 0.122$\pm$0.005 & 10.8$\pm$0.1 & 
				  11.48$\pm$0.86 & 5.80$\pm$0.09 &   \\
Lynds 495  & 12.18$\pm$0.10 & 0.26$\pm$0.02 & 0.22$\pm$0.02 & 12.6$\pm$0.6 & 
                  0.60$\pm$0.06 & 10 & C$^{18}$O ($J=1-0$) \\
				  & 12.18$\pm$0.01 & 0.26$\pm$0.01 & 0.23$\pm$0.02 & 12.7$\pm$0.4 & 
				  0.54$\pm$0.03 & 10 &   \\
Lynds 498  & 12.51$\pm$0.01 & 0.18$\pm$0.03 & 0.14$\pm$0.04 & 11.0$\pm$0.6 & 
                  0.41$\pm$0.07 & 10 & C$^{18}$O ($J=1-0$) \\
				  & 12.48$\pm$0.01 & 0.29$\pm$0.02 & 0.27$\pm$0.03 & 13.6$\pm$0.7 & 
				  0.31$\pm$0.02 & 10 &   \\
\tableline
\end{tabular}
\tablenotetext{a}{Observed FWHM line width.}
\tablenotetext{b}{Non-thermal line width. We assumed thermal line width $\Delta V_{\rm T}$ is equal to 0.443 km s${}^{-1}$ ($T_{\rm k}=10$ K).}
\tablenotetext{c}{Temperature including non-thermal (turbulent) contribution ($T_{\rm eff}=T_{\rm k}+T_{\rm NT}$). }
\tablenotetext{d}{Peak optical depth. The value for FeSt 1-457 is the sum of the peak optical depths of the seven hyperfine components.}
\tablenotetext{e}{We assumed the excitation temperature for the C$^{18}$O spectra to be 10 K.}
\tablenotetext{f}{The first row refers to the values from the spectrum toward the center of each globule, and the second row refers to the values from the composite spectrum of the central nine positions. The error represents $1 \sigma$ error in the line profile fitting.}
\end{center}
\end{table}
\end{landscape}

\clearpage
\begin{landscape}
\begin{table}
\tablewidth{1200pt}
\begin{center}
\caption{Physical properties of globules I}
\begin{tabular}{lccllcccclclc}
\tableline\tableline
Name & 
\multicolumn{1}{c}{R.A. (J2000)\tablenotemark{a}} & 
\multicolumn{1}{c}{Dec. (J2000)\tablenotemark{a}} & 
\multicolumn{1}{c}{$\theta {}_{R}$} & 
\multicolumn{1}{c}{$\xi {}_{\rm max}$} & 
\multicolumn{1}{c}{${n}_{\rm c} / {n}_{\rm edge}$\tablenotemark{b}} & 
\multicolumn{1}{c}{$A_{V}$\tablenotemark{c}} & 
\multicolumn{1}{c}{$e$\tablenotemark{d}} & 
\multicolumn{1}{c}{P.A.\tablenotemark{d}} & 
\multicolumn{1}{c}{$r_{\rm fit}$\tablenotemark{e}} & 
\multicolumn{1}{c}{$\chi {}^{2}$} & 
\multicolumn{1}{c}{Stability\tablenotemark{f}} & 
Reference\tablenotemark{g} 
\\
  & 
\multicolumn{1}{c}{($^{\rm h}$ $^{\rm m}$ $^{\rm s}$)} & 
\multicolumn{1}{c}{($^{\circ}$ $^{'}$ $^{''}$)} & 
\multicolumn{1}{c}{($''$)} & 
  & 
  & 
\multicolumn{1}{c}{(mag)} & 
  & 
\multicolumn{1}{c}{($^{\circ}$)} & 
\multicolumn{1}{c}{($''$)} & 
  & 
  & 
  
\\
\tableline
CB 87          & 17 24 58.1 & $-$24 06 49 & 87.5$\pm$2.6   & 5.1$\pm$0.4  & 8.1   & 6.6   & 0.52  & 120    & 
                    $<$110   & 3.19  & Stable & A\\
CB 110         & 18 05 54.7 & $-$18 25 02 & 61.1$\pm$3.1   & 14.0$\pm$3.0 & 94.3  & 19.7  & ----- & -----  & 
                    $<$70    & 1.17  & Unstable & A\\
CB 131         & 18 17 00.5 & $-$18 02 05 & 103$\pm$7.4   & 16.3$\pm$5.1 & 139   & 32.9  & ----- & -----  & 
                    $<$150   & 3.22  & Unstable & A\\
CB 134         & 18 22 46.3 & $-$01 42 50 & 59.6$\pm$4.3   & 18.5$\pm$4.9 & 187   & 26.8  & ----- & -----  & 
                    $<$55    & 0.44  & Unstable & A\\
CB 161         & 18 53 56.0 & $-$07 26 07  & 62.5$\pm$6.9   & 8.1$\pm$1.4  & 25.1  & 13.0  & 0.72  & 125    & 
                    $<$50    & 0.11  & Unstable & A\\
CB 184         & 19 13 51.7 & $+$16 27 27 & 112$\pm$20     & 8.1$\pm$1.6  & 24.9  & 11.2  & 0.68  & 0      & 
                    $<$75    & 0.15  & Unstable & A\\
CB 188         & 19 20 15.7 & $+$11 36 07 & 127$\pm$19     & 16.0$\pm$2.9 & 132   & 27.2  & ----- & -----  & 
                    $<$100   & 5.17  & Unstable & A\\
FeSt 1-457     & 17 35 47.5 & $-$25 32 59 & 144$\pm$11     & 12.6$\pm$2.0 & 74.5  & 41.0  & ----- & -----  & 
                    $<$140   & 1.41  & Unstable & A\\
Lynds 495      & 18 38 57.4 & $-$06 44 06 & 75.0$\pm$7.8   & 7.2$\pm$1.4  & 18.6  & 12.7  & ----- & -----  & 
                    $<$65    & 0.14  & Unstable & A\\
Lynds 498      & 18 40 10.5 & $-$06 40 43 & 75.0$\pm$3.0   & 4.7$\pm$0.4  & 6.63  & 10.9  & ----- & -----  & 
                    $<$70    & 0.84  & Stable & A\\
\tableline
Barnard 68     & --------   & --------      & 100            & 6.9$\pm$0.2  & 16.6  & ----- & ----- & -----  & 
                     -----    & ----- & Unstable & B\\
Barnard 335   & 19 37 00.9 & $+$07 34 10.0 & 125            & 12.5$\pm$2.6 & 73.1  & ----- & ----- & -----  & 
                     $<$100   & 2.93  & Unstable & C\\
Coalsack        & 12 31 38.6 & $-$63 43 42.5 & 290            & 5.8          & 10.9  & 11.5  & ----- & -----  & 
                     55$-$400 & ----- & Stable   & D\\
                     & --------   & --------      & 140            & 7.0$\pm$0.3  & 17.2  & 23.0  & ----- & -----  & 
                     -----    & ----- & Unstable & E\\
Lynds 694-2   & 19 41 04.4 & $+$10 57 00.9 & 54             & 25$\pm$3     & 364   & ----- & ----- & -----  & 
                     $<$83    & 1.12  & Unstable & F\\
\tableline
\end{tabular}
\tablenotetext{a}{Position of core center (centroid of $A_V$ distribution).}
\tablenotetext{b}{Center-to-edge density contrast determined from $\xi {}_{\rm max}$ value.}
\tablenotetext{c}{$A_{V}$ at core center.}
\tablenotetext{d}{Fitted ellipticity and position angle.}
\tablenotetext{e}{Profile fitting region in the Bonnor-Ebert fit.}
\tablenotetext{f}{Stability of equilibrium state of the Bonnor-Ebert sphere.}
\tablenotetext{g}{ 
(A) This work; (B) Alves et al. (2001); (C) Harvey et al. (2001); (D) Lada et al. (2004); 
(E) Racca et al. (2002); (F) Harvey et al. (2003)
}
\end{center}
\end{table}
\end{landscape}

\clearpage
\begin{table}
\tablewidth{1000pt}
\begin{center}
\caption{Physical properties of globules II}
\begin{tabular}{lllllllc}
\tableline \tableline
Name & 
\multicolumn{1}{c}{$T$\tablenotemark{a}} & 
\multicolumn{1}{c}{$D$\tablenotemark{b}} & 
\multicolumn{1}{c}{$R$\tablenotemark{c}} & 
\multicolumn{1}{c}{$n {}_{\rm c}$\tablenotemark{c}} & 
\multicolumn{1}{c}{$M$\tablenotemark{c}} & 
\multicolumn{1}{c}{$P_{\rm ext}$\tablenotemark{c}} & 
Reference\tablenotemark{d} 
\\
  & 
\multicolumn{1}{c}{(K)} & 
\multicolumn{1}{c}{(pc)} & 
\multicolumn{1}{c}{(AU)} & 
\multicolumn{1}{c}{(cm${}^{-3}$)} & 
\multicolumn{1}{c}{(M${}_{\odot}$)} & 
\multicolumn{1}{c}{(K cm${}^{-3}$)} & 
\\
\tableline
CB87       & 6.0$\pm$0.4   & 160 & (1.40$\pm$0.04)$\times 10^{4}$ & (3.2$\pm$0.4)$\times 10^{4}$ & 
           0.76$\pm$0.07 & (2.9$\pm$0.4)$\times 10^{4}$ & A \\
           & 11.4          & 304 & (2.66$\pm$0.08)$\times 10^{4}$ & (1.7$\pm$0.2)$\times 10^{4}$ & 
		   2.73$\pm$0.24 & (2.9$\pm$0.4)$\times 10^{4}$ & A \\
CB 110     & 7.0$\pm$0.4   & 180 & (1.10$\pm$0.06)$\times 10^{4}$ & (4.5$\pm$1.7)$\times 10^{5}$ & 
           0.74$\pm$0.17 & (4.0$\pm$0.9)$\times 10^{4}$ & A \\
           & 21.8          & 561 & (3.43$\pm$0.18)$\times 10^{4}$ & (1.5$\pm$0.6)$\times 10^{5}$ & 
		   7.21$\pm$1.64 & (4.0$\pm$0.9)$\times 10^{4}$ & A \\
CB 131     & 14.0$\pm$0.8  & 180 & (1.85$\pm$0.13)$\times 10^{4}$ & (4.4$\pm$2.3)$\times 10^{5}$ & 
           2.43$\pm$0.73 & (5.3$\pm$1.5)$\times 10^{4}$ & A \\
           & 25.1          & 323 & (3.32$\pm$0.24)$\times 10^{4}$ & (2.5$\pm$1.3)$\times 10^{5}$ & 
		   7.83$\pm$2.35 & (5.3$\pm$1.5)$\times 10^{4}$ & A \\
CB 134     & 13.0$\pm$0.8  & 260 & (1.55$\pm$0.11)$\times 10^{4}$ & (7.6$\pm$3.4)$\times 10^{5}$ & 
           1.85$\pm$0.50 & (6.3$\pm$1.7)$\times 10^{4}$ & A \\
           & 13.2          & 264 & (1.57$\pm$0.11)$\times 10^{4}$ & (7.5$\pm$3.3)$\times 10^{5}$ & 
		   1.91$\pm$0.52 & (6.3$\pm$1.7)$\times 10^{4}$ & A \\
CB 161     & 14.0$\pm$1.1  & 400 & (2.50$\pm$0.28)$\times 10^{4}$ & (6.1$\pm$1.4)$\times 10^{4}$ & 
           3.51$\pm$0.90 & (4.0$\pm$1.5)$\times 10^{4}$ & A \\
           & 12.5         & 357 & (2.23$\pm$0.25)$\times 10^{4}$ & (7.0$\pm$1.6)$\times 10^{4}$ & 
		   2.79$\pm$0.72 & (4.0$\pm$1.5)$\times 10^{4}$ & A \\
CB 184     & 14.0$\pm$1.0  & 300 & (3.35$\pm$0.61)$\times 10^{4}$ & (3.4$\pm$0.5)$\times 10^{4}$ & 
           4.70$\pm$1.76 & (2.2$\pm$1.6)$\times 10^{4}$ & A \\
           & 15.5          & 332 & (3.71$\pm$0.67)$\times 10^{4}$ & (3.0$\pm$0.4)$\times 10^{4}$ & 
		   5.76$\pm$2.17 & (2.2$\pm$1.6)$\times 10^{4}$ & A \\
CB 188     & 18.0$\pm$1.1  & 300 & (3.80$\pm$0.57)$\times 10^{4}$ & (1.3$\pm$0.3)$\times 10^{5}$ & 
           6.45$\pm$2.05 & (2.1$\pm$1.1)$\times 10^{4}$ & A \\
           & 19.0          & 317 & (4.01$\pm$0.60)$\times 10^{4}$ & (1.2$\pm$0.2)$\times 10^{5}$ & 
		   7.19$\pm$2.28 & (2.1$\pm$1.1)$\times 10^{4}$ & A \\
FeSt 1-457 & 24.0$\pm$1.7  & 160 & (2.30$\pm$0.18)$\times 10^{4}$ & (2.9$\pm$0.8)$\times 10^{5}$ & 
           5.41$\pm$1.13 & (1.1$\pm$0.3)$\times 10^{5}$ & A \\
           & 10.9          & 73  & (1.04$\pm$0.08)$\times 10^{4}$ & (6.5$\pm$1.7)$\times 10^{5}$ & 
		   1.12$\pm$0.23 & (1.1$\pm$0.3)$\times 10^{5}$ & A \\
Lynds 495  & 8.0$\pm$0.7   & 200 & (1.50$\pm$0.16)$\times 10^{4}$ & (7.6$\pm$2.1)$\times 10^{4}$ & 
           1.19$\pm$0.31 & (3.9$\pm$1.4)$\times 10^{4}$ & A \\
           & 12.6          & 315 & (2.36$\pm$0.25)$\times 10^{4}$ & (4.8$\pm$1.4)$\times 10^{4}$ & 
		   2.95$\pm$0.77 & (3.9$\pm$1.4)$\times 10^{4}$ & A \\
Lynds 498  & 11.0$\pm$0.9  & 200 & (1.50$\pm$0.06)$\times 10^{4}$ & (4.3$\pm$0.5)$\times 10^{4}$ & 
           1.42$\pm$0.16 & (8.5$\pm$1.7)$\times 10^{4}$ & A \\
           & 11.0          & 200 & (1.50$\pm$0.06)$\times 10^{4}$ & (4.3$\pm$0.5)$\times 10^{4}$ & 
		   1.42$\pm$0.16 & (8.5$\pm$1.7)$\times 10^{4}$ & A \\
\tableline
Barnard 68   & 16            & 125 & 1.25$\times 10^{4}$            & --------                       & 
                    2.10          & 1.8$\times 10^{5}$         & B \\
                    & 10$\pm$1.2    &  85 & 0.85$\times 10^{4}$            & --------                       & 
		            0.90          & 1.7$\times 10^{5}$         & G \\
Barnard 335  & --------      & 250 & 3.125$\times 10^{4}$           & 3.0$\times 10^{4}$             & 
                    14.0          & --------                   & C \\
Coalsack       & 19            & 150 & 4.35$\times 10^{4}$            & 1.8$\times 10^{4}$             & 
                    6.1$\pm$0.5   & --------                   & D \\
                    & 15            & 180 & 2.52$\times 10^{4}$            & 5.4$\times 10^{4}$             & 
				    4.50          & 6.5$\times 10^{4}$         & E \\
Lynds 694-2 & --------      & 250 & 1.35$\times 10^{4}$            & 3.1$\times 10^{5}$             & 
                    3.00          & 1.1$\times 10^{5}$         & F \\
\tableline
\end{tabular}
\tablenotetext{a}{First row: value from the Bonnor-Ebert fit using assumed distance to the globules (see Table 1), Second row: value from molecular line observations ($T_{\rm eff}$; see, Table 3).
}
\tablenotetext{b}{ 
First row: same value as listed in Table 1, Second row: calibrated value by using $T_{\rm eff}$ to scale the fitted Bonnor-Ebert model parameters (see, \S 4.2.2).
}
\tablenotetext{c}{
The values in the first and second rows are derived using the distances listed in the third column.
}
\tablenotetext{d}{ 
(A) This work; 
(B)-(F) Same references as those in Table 4; 
(G) Hotzel et al. (2002a). 
}
\end{center}
\end{table}


\clearpage
\begin{figure}
\epsscale{0.9}
\plotone{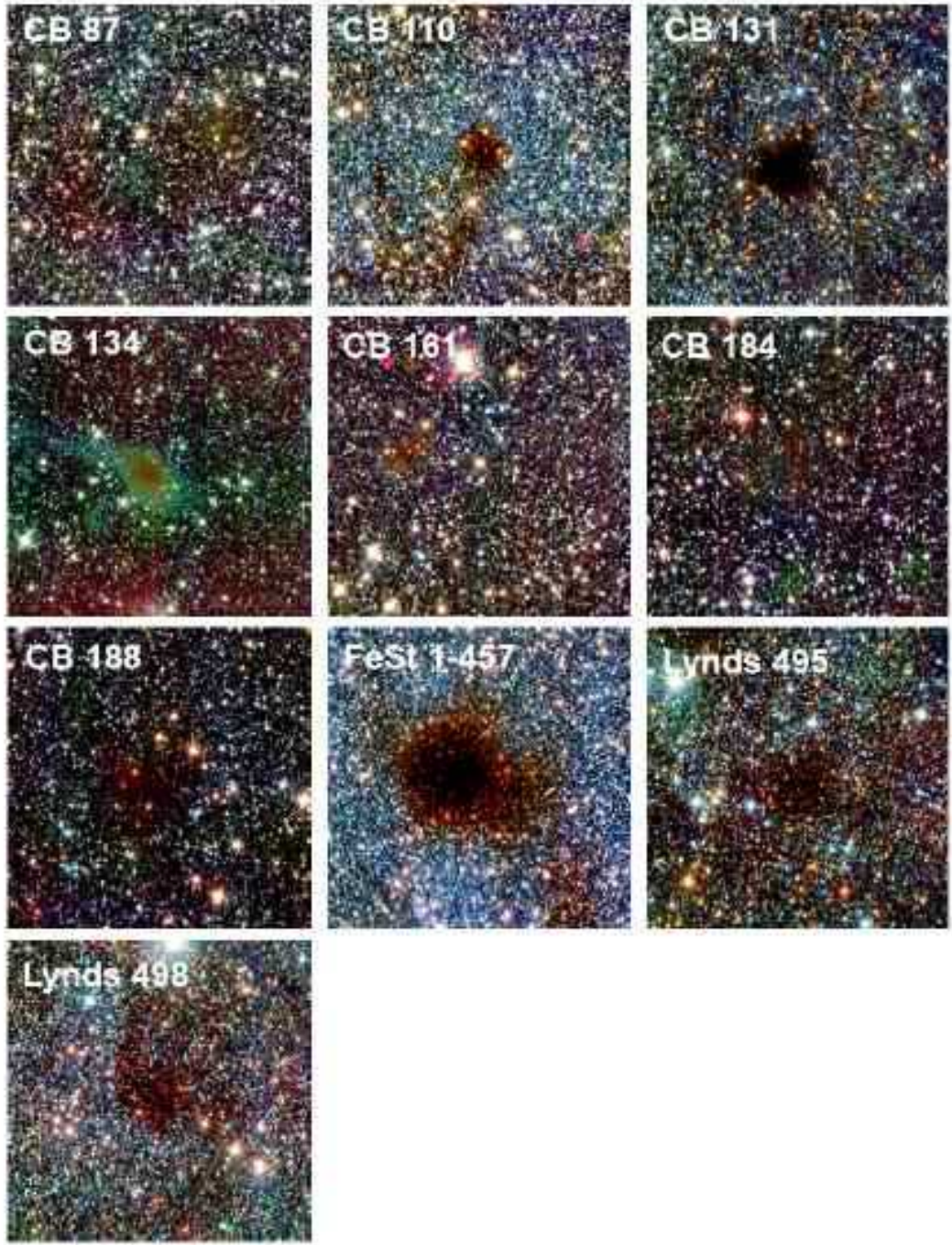}
\caption{$J$, $H$, and  $K_{s}$ three-color composite images for ten Bok globules ($J$: blue, $H$: green,  $K_{s}$: red). Each image size is 930$\times$930 pixel ($\sim$7 arcminutes). North is up, and east is toward the left.}
\end{figure}

\clearpage
\begin{figure}
\epsscale{0.9}
\plotone{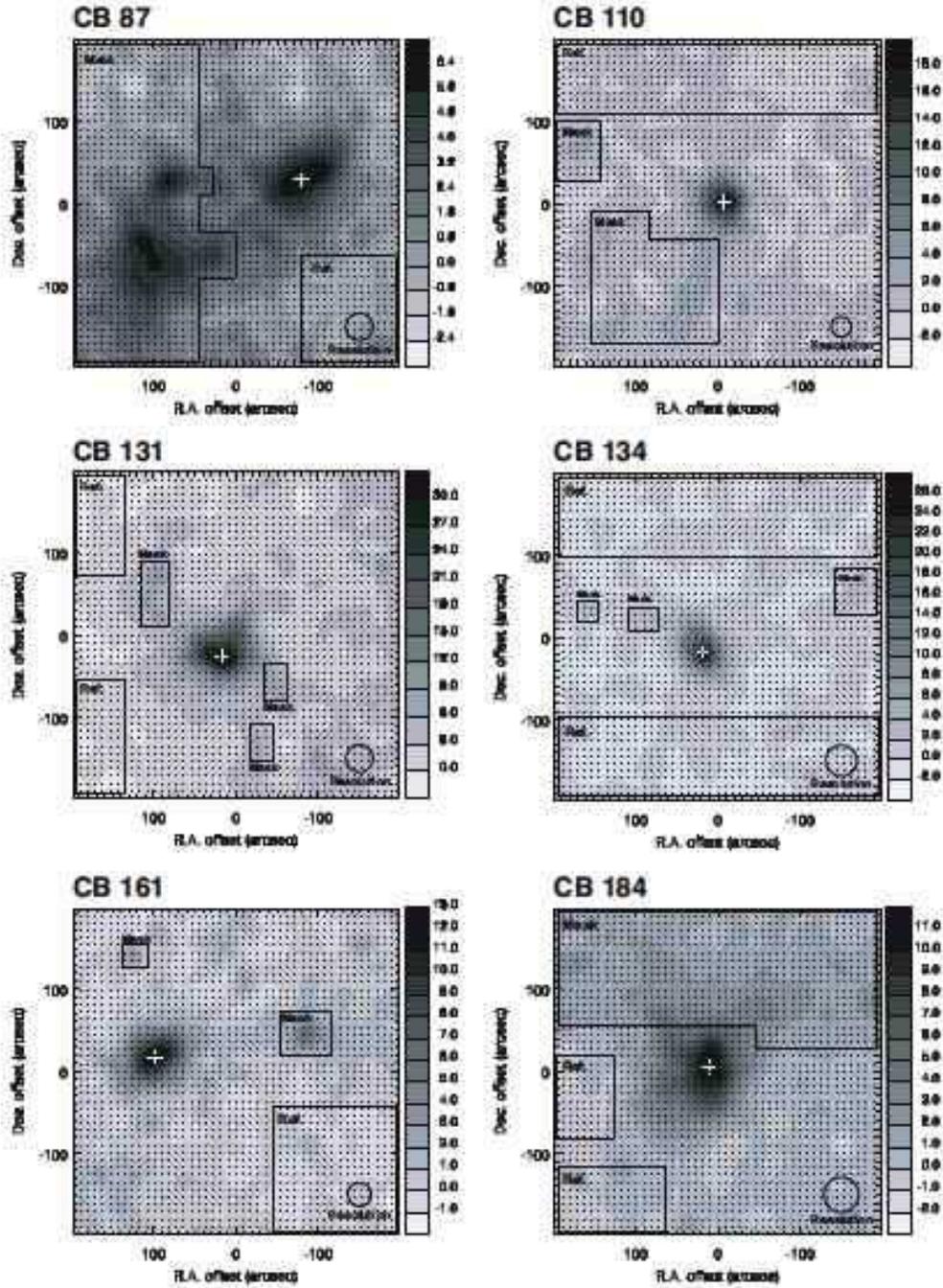}
\caption{$A_{V}$ distribution for Bok globules. The area enclosed by the broken line is the reference field expected to be free from dust extinction, and the area enclosed by the dotted line is the masked region which is excluded in the derivation of the radial $A_{V}$ profile. }
\end{figure}

\clearpage
\setcounter{figure}{1}
\begin{figure}
\epsscale{0.9}
\plotone{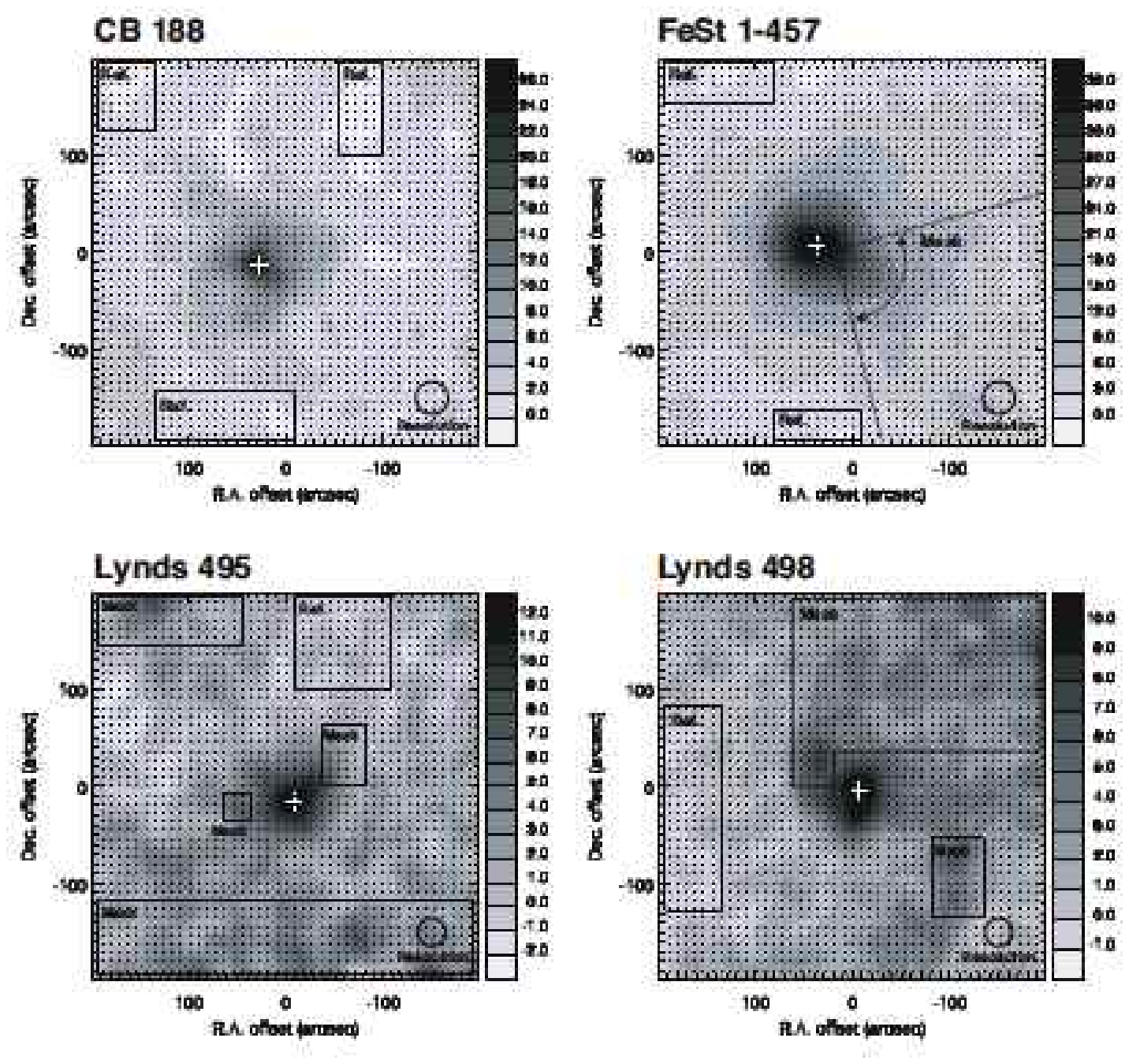}
\caption{{\it continued.}}
\end{figure}

\clearpage
\begin{figure}
\plotone{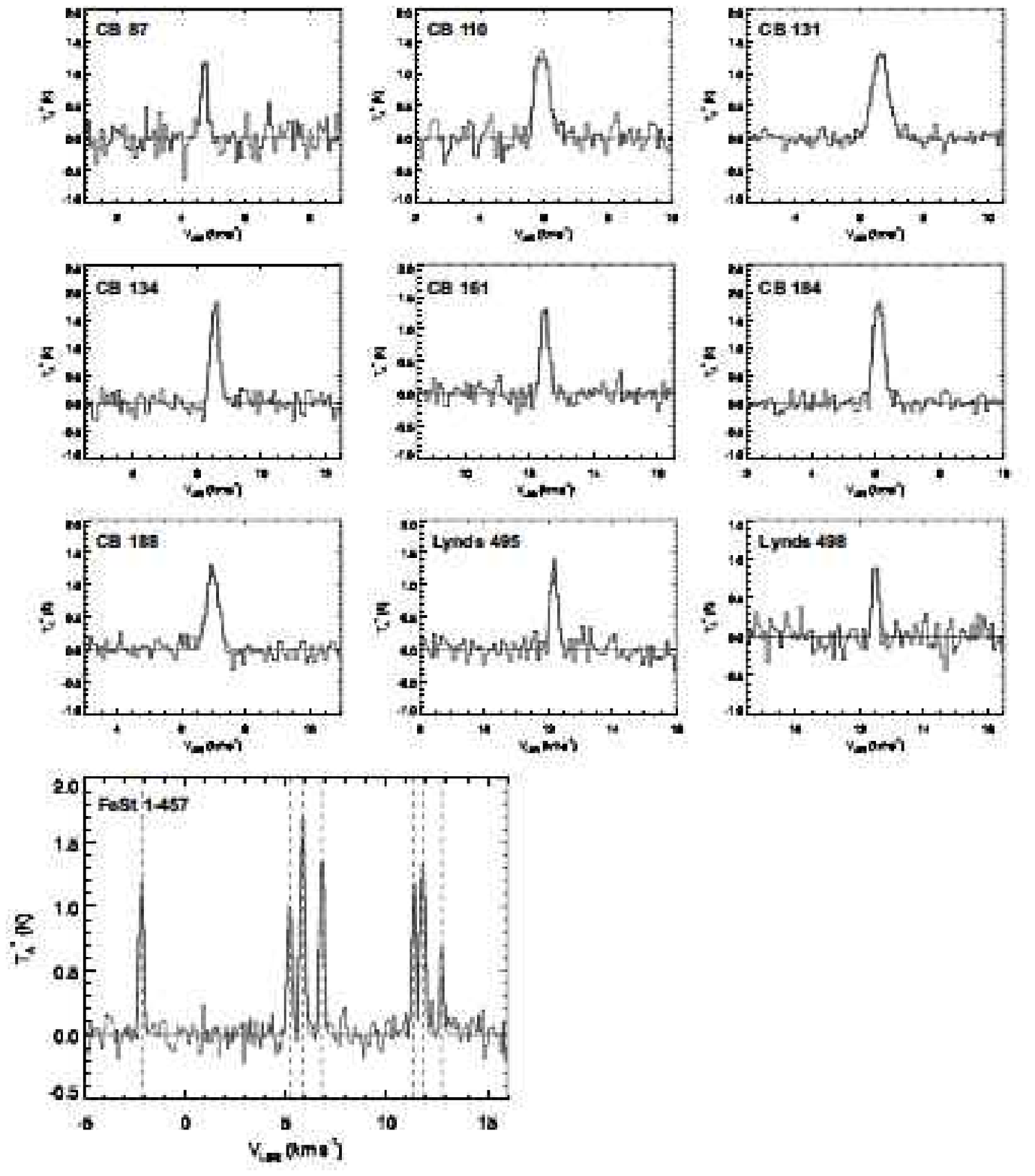}
\epsscale{1}
\caption{Line profiles toward the center of ten globules 
(N${}_{2}$H${}^{+}$ for FeSt 1-457, C${}^{18}$O for the other globules). 
The gray solid line denotes fitting result of each spectrum.}
\end{figure}

\clearpage
\begin{figure}
\epsscale{0.8}
\plotone{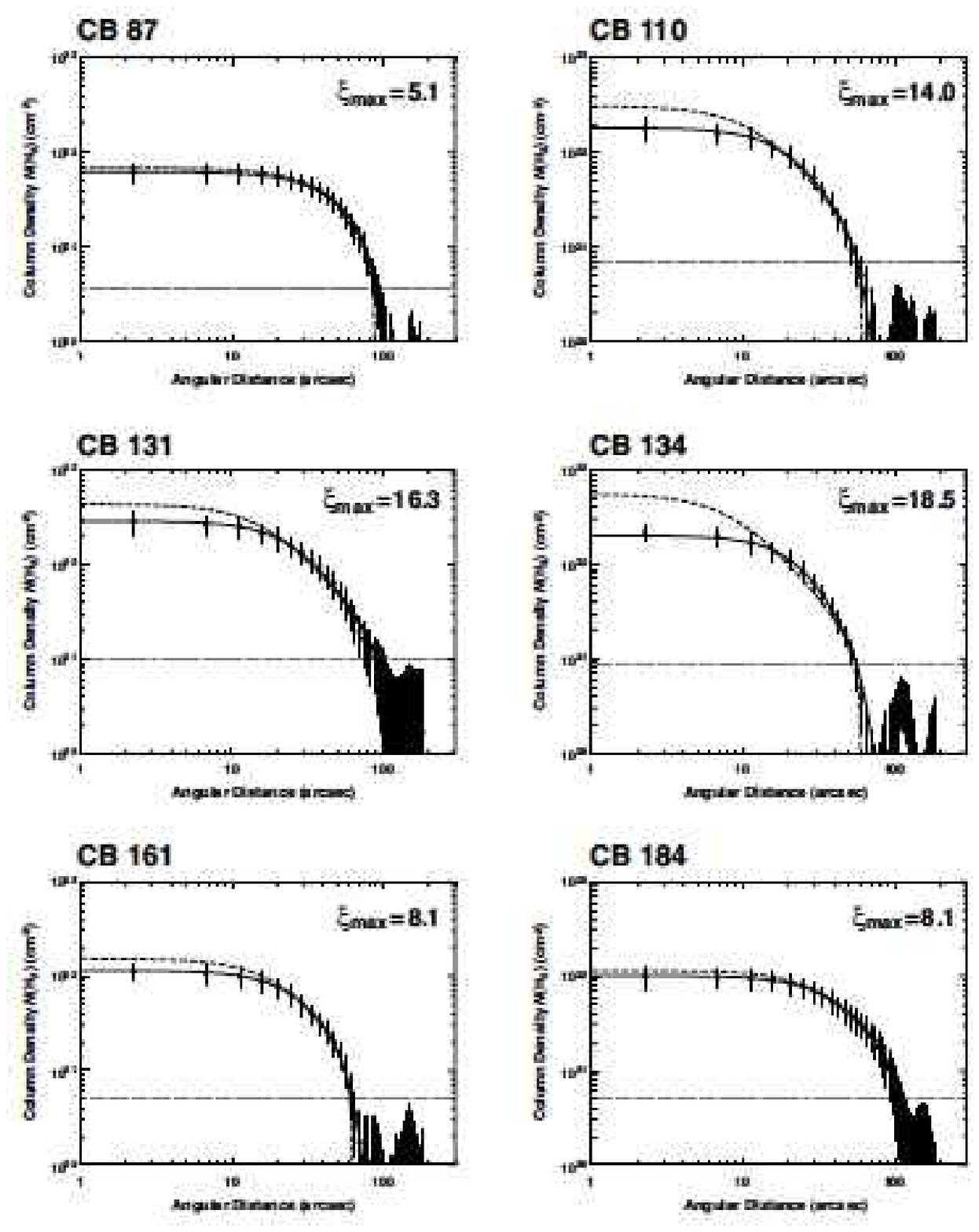}
\caption{Radial column density profile and the Bonnor-Ebert model fit for the ten Bok globules. The dots and error bars represent the average $N$(H${}_{2}$) values at each annulus arranged at intervals of 9$''$ and the rms dispersion of data points in each annulus, respectively. The solid line denotes the best-fit Bonnor-Ebert profile which is convolved to the observing resolution (typically $\sim$30$''$) with the same beam as used in the $A_{V}$ measurements. The dashed line denotes the Bonnor-Ebert profile before the convolution. The gray dot-dashed line denotes the 1 $\sigma$ deviation of the column density measured in the reference field.}
\end{figure}

\clearpage
\setcounter{figure}{3}
\begin{figure}
\epsscale{0.8}
\plotone{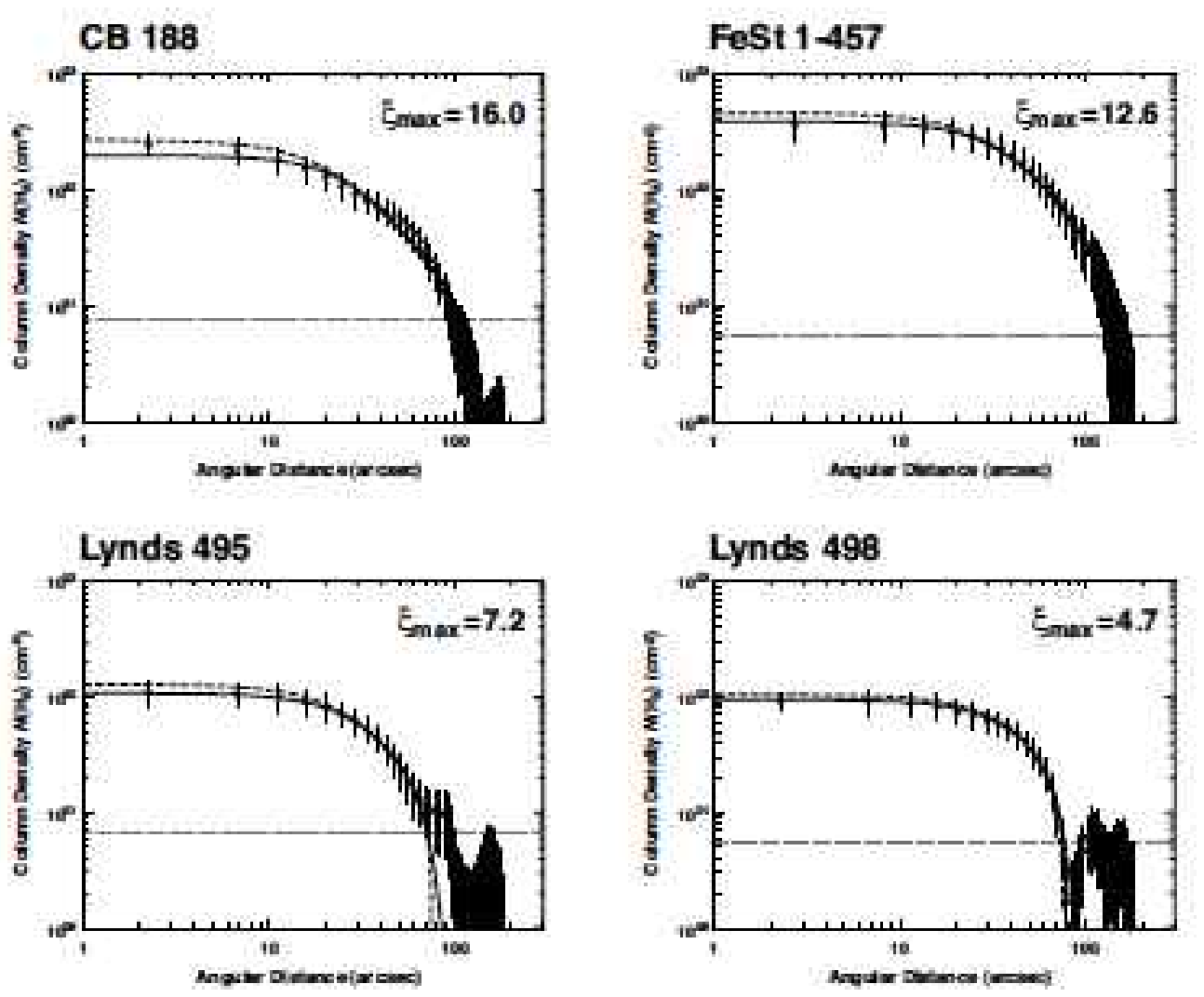}
\caption{{\it continued.}}
\end{figure}

\clearpage
\begin{figure}
\epsscale{0.9}
\plotone{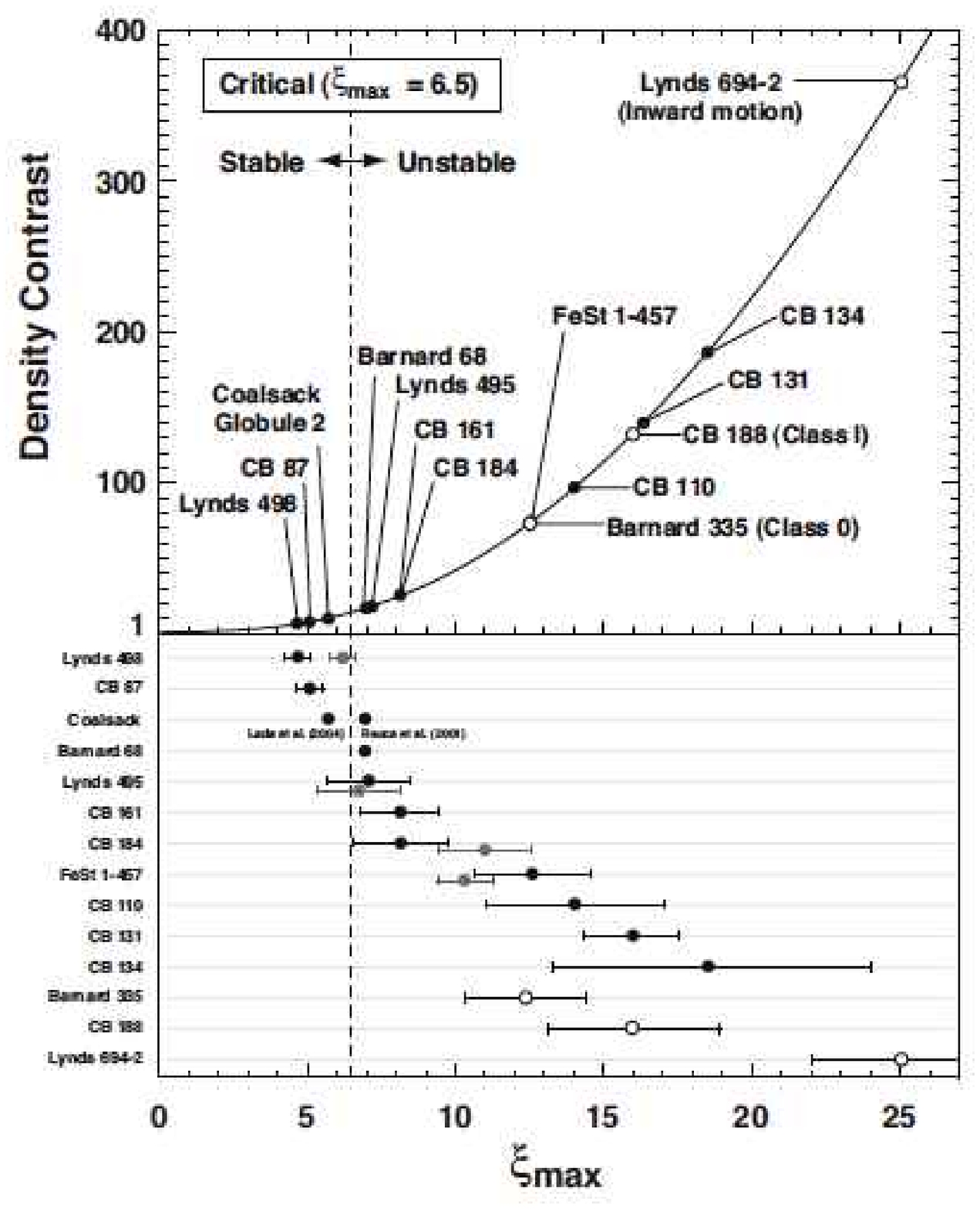}
\caption{The solid line shows the relation between $\xi {}_{{\rm max}}$ and density contrast (center-to-edge density ratio). The location of our 10 globules and previously reported 4 globules (Barnard 335, Harvey et al. 2001; Barnard 68, Alves et al. 2001; Coalsack Globule II, Racca et al. 2002; Lada et al. 2004; Lynds 694-2, Harvey et al. 2003) are shown in the figure. Filled and open black circles denote starless and star-forming globules, respectively. Filled gray circles denote the fitting results when the masked region on the $A_{V}$ map (see Fig. 2) is included in the derivation of the column density profile. The vertical dashed line denotes critical state of $\xi {}_{{\rm max}}$$=6.5$. In the lower panel of the figure, horizontal error bars denote 1$\sigma$ error in the $\chi ^{2}$ density profile fitting.}
\end{figure}

\clearpage
\begin{figure}
\epsscale{0.5}
\plotone{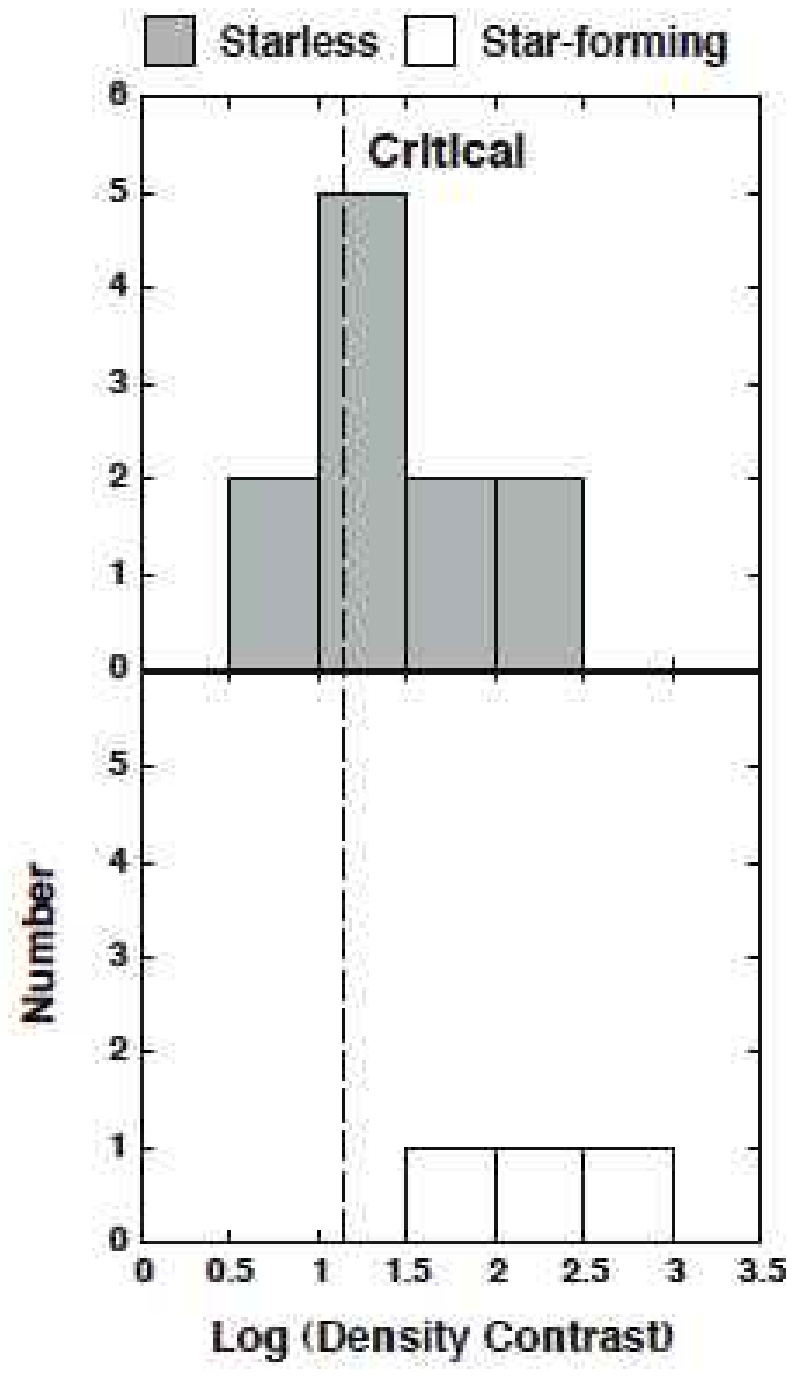}
\caption{Histograms of logarithmic density contrast for starless and star-forming globules. The vertical broken line denotes the density contrast of the critical Bonnor-Ebert sphere.}
\end{figure}

\clearpage
\begin{figure}
\epsscale{0.8}
\plotone{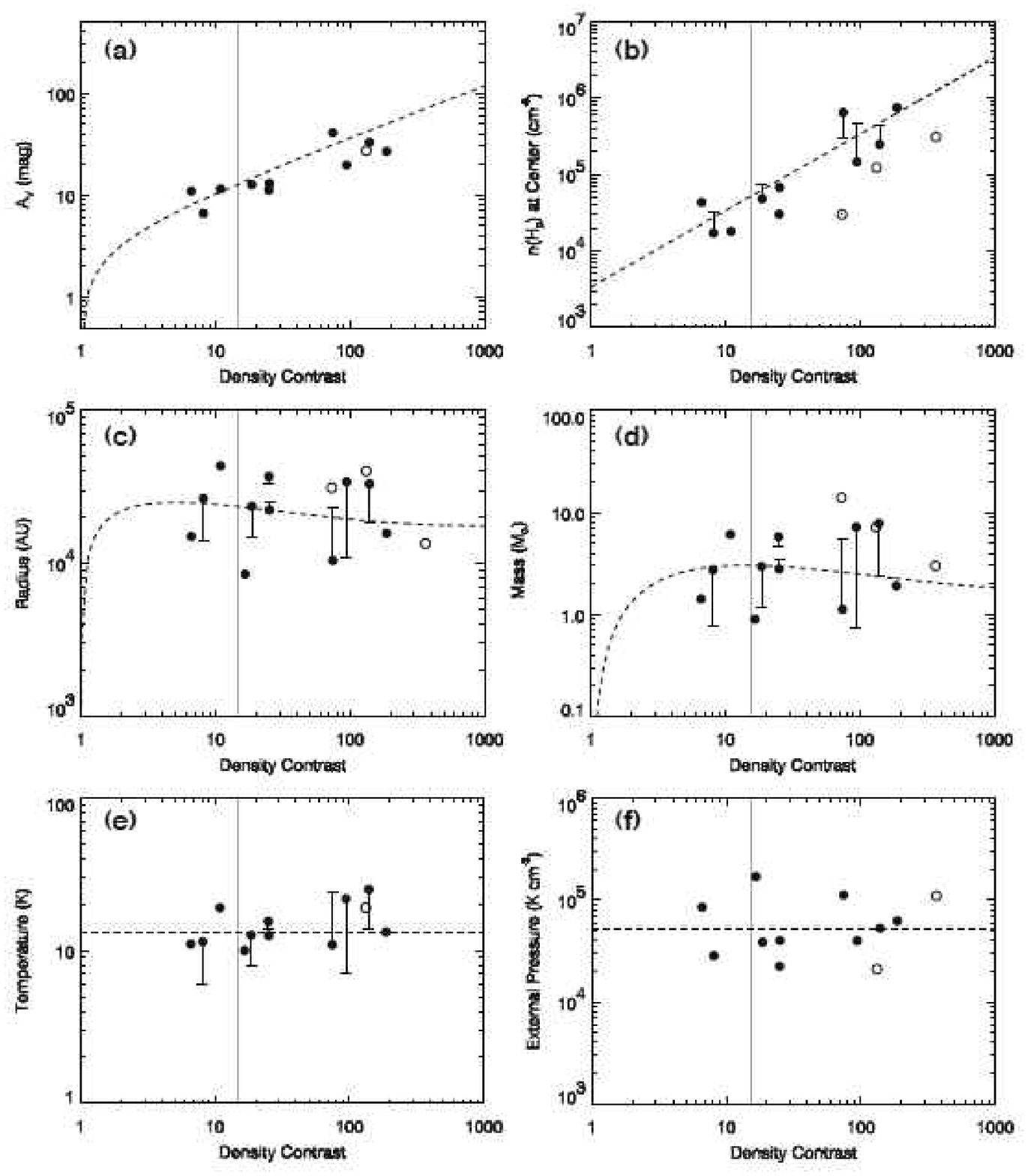}
\caption{Correlations between density contrast and the other physical parameters. The panels of (a) to (f) show the correlation diagrams for $A_{V}$, central density, radius, mass, temperature, and external pressure, respectively. Filled and open circles denote starless and star-forming globules, respectively. The vertical gray line denotes the density contrast of the critical Bonnor-Ebert sphere. The plotted data points are the physical quantities after distance-correction (see, \S 4.2.2, Table 5). The physical quantities without distance-correction are shown as the symbol $\lq \lq${\boldmath $-$}" in each panel. The broken lines represent the relationships for the Bonnor-Ebert spheres with a constant temperature $T_{\rm eff}$ of 13.2 K and $P_{\rm ext}$ of $5.3 \times 10^4$ K cm$^{-3}$. 
}
\end{figure}

\clearpage
\begin{figure}
\epsscale{1}
\plotone{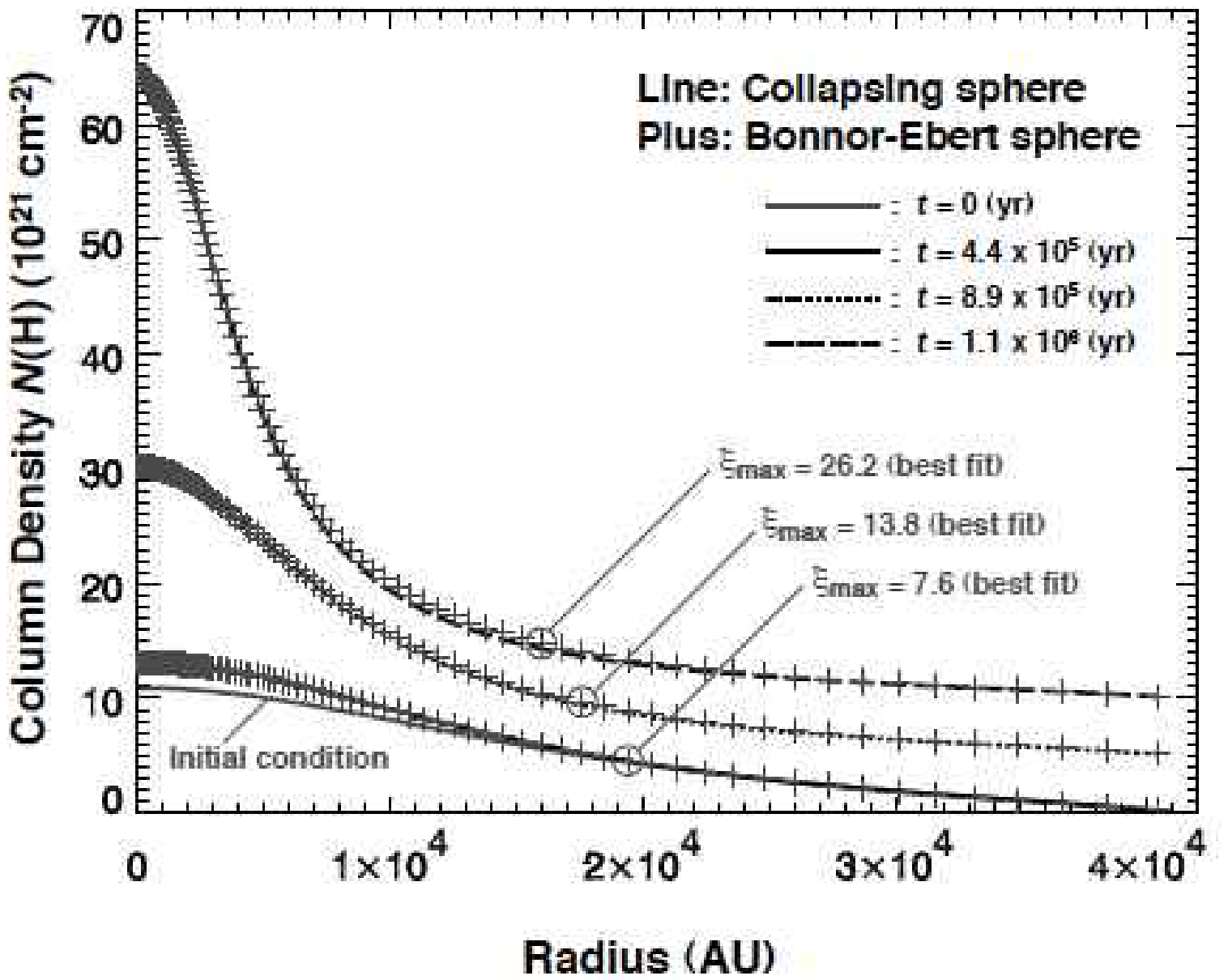}
\caption{
Radial column density distribution of a collapsing gas sphere at some specific times (black solid, dotted, and broken line) started from the initial condition of a nearly critical Bonnor-Ebert density structure (gray solid line). 
Dotted and broken lines are shifted by 5 and 10 ($10^{21}$ cm${}^{-2}$) in the figure, respectively. Best-fit solution of the Bonnor-Ebert sphere for each line is also plotted as plus symbol.
}
\end{figure}

\clearpage
\begin{figure}
\epsscale{1}
\plotone{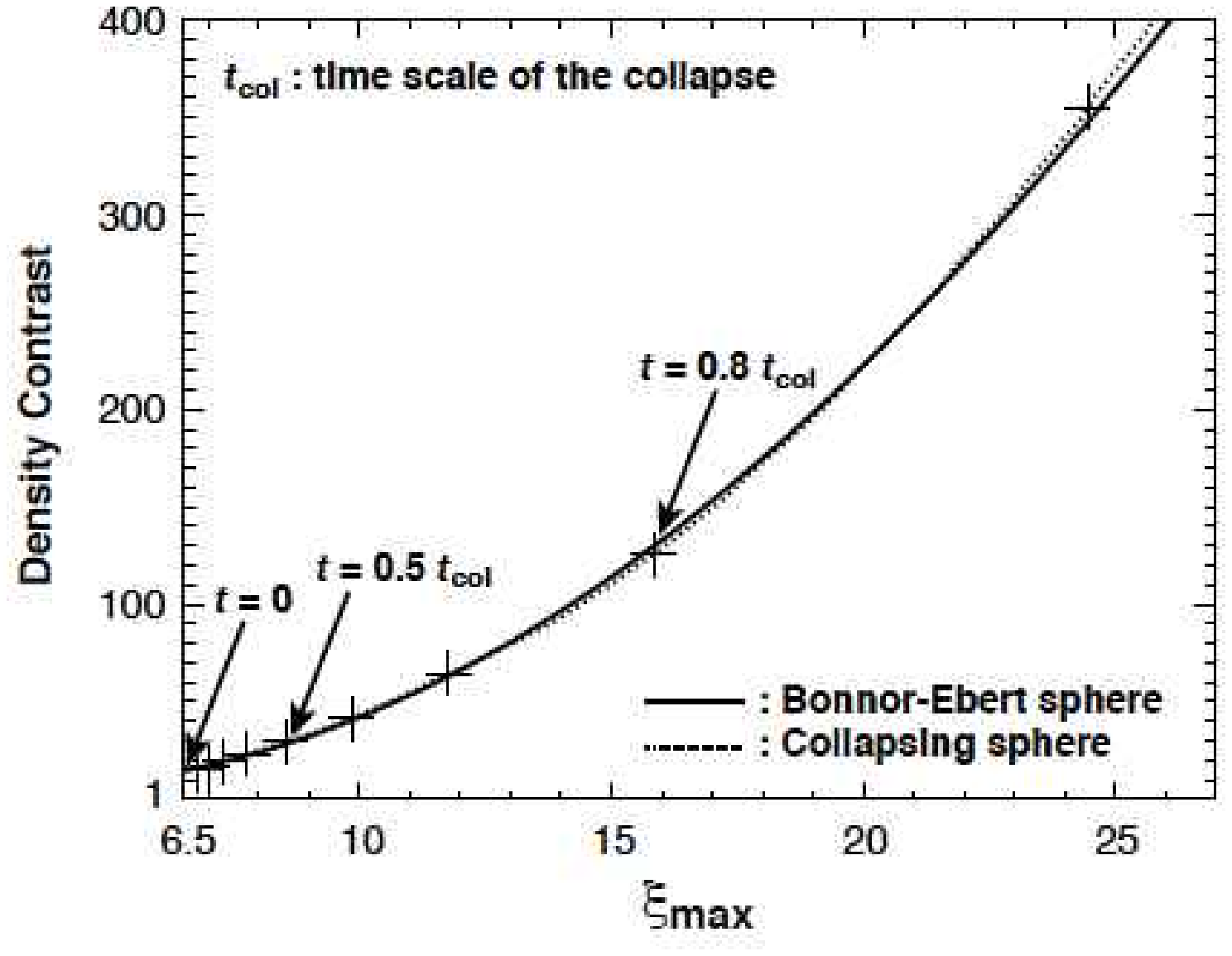}
\caption{
The solid line shows the relation between $\xi {}_{{\rm max}}$ and density contrast for the Bonnor-Ebert sphere (same as Fig. 5). The dotted line denotes the similar relation for a collapsing sphere starting from a nearly critical Bonnor-Ebert sphere ($\alpha = 1.1$ model; Aikawa et al. 2004); the column density profiles of the collapsing sphere were fitted with static Bonnor-Ebert sphere solutions, and the best-fit $\xi {}_{{\rm max}}$ values are plotted against the density contrast values of the collapsing sphere. The plus symbols denote the elapsed times from the onset of collapse at 10\% intervals of the total collapse time ($t_{\rm col}$). 
}
\end{figure}

\clearpage
\begin{figure}
\epsscale{0.8}
\plotone{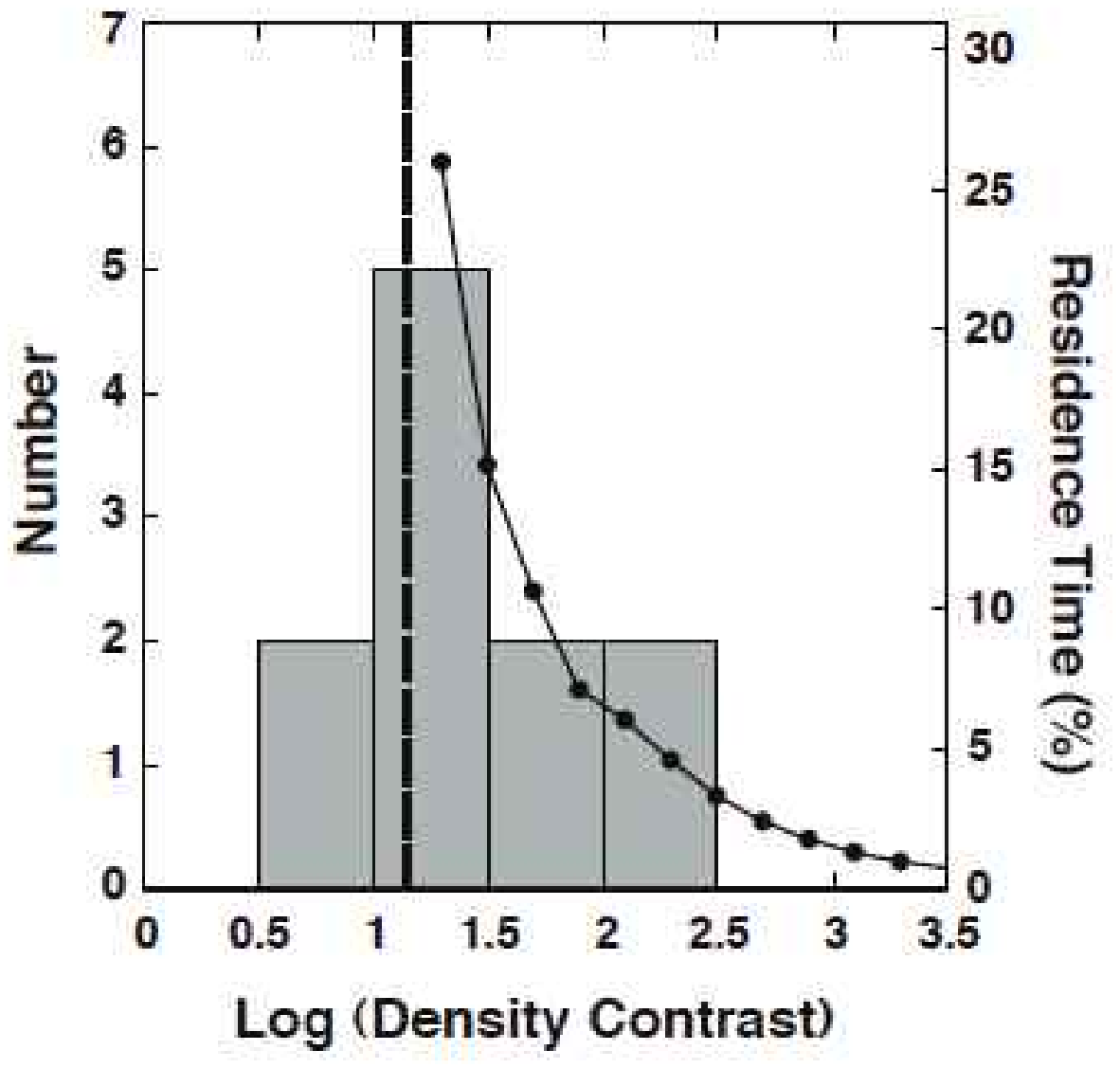}
\caption{The histogram of the logarithmic density contrast for starless globules (the same as Figure 6). The solid line with dots shows the time evolution of the model collapsing sphere; each dot represents the residence time of a collapsing sphere measured at 0.2 intervals in the logarithmic density contrast. The vertical axis on the right side shows the percentage of the residence time with respect to the total collapse time.}
\end{figure}

\end{document}